\documentclass[fleqn,usenatbib,useAMS]{mnras}

\usepackage{graphicx}	
\usepackage{amsmath}	
\usepackage{amssymb}	
\usepackage{multicol}        
\usepackage{bm}		
\usepackage{pdflscape}	

\usepackage{subcaption}
\usepackage{caption}

\newcommand{\red}[1]{\textcolor{black}{#1}}
\newcommand{\mage}[1]{\textcolor{black}{#1}}

\newcommand{\ahod}{\textsc{AbacusHOD}}

\usepackage[T1]{fontenc}
\usepackage{ae,aecompl}

\usepackage{newtxtext,newtxmath}

\title[RSD Emulator]{Stringent $\sigma_8$ constraints from small-scale galaxy clustering using a hybrid MCMC+emulator framework}

\author[S. Yuan et al.]{
Sihan Yuan,$^{1, 2, 3}$\thanks{E-mail: sihany@stanford.edu}
Lehman H. Garrison, $^{4}$
Daniel J. Eisenstein, $^{3}$
and Risa H. Wechsler $^{1, 2, 5}$
\\
$^{1}$Kavli Institute for Particle Astrophysics and Cosmology, 452 Lomita Mall, Stanford University, Stanford, CA 94305, USA\\
$^{2}$SLAC National Accelerator Laboratory, 2575 Sand Hill Road, Menlo Park, CA 94025, USA\\
$^{3}$Center for Astrophysics | Harvard \& Smithsonian, 60 Garden St., Cambridge, MA 02138, USA\\
$^{4}$Center for Computational Astrophysics, Flatiron Institute, 162 Fifth Avenue, New York, NY 10010, USA\\
$^{5}$Department of Physics, Stanford University, 382 Via Pueblo Mall, Stanford, CA 94305, USA
}

\date{Last updated 2020 June 10; in original form 2013 September 5}

\pubyear{2020}

\begin{document}
\label{firstpage}
\pagerange{\pageref{firstpage}--\pageref{lastpage}}
\maketitle

\begin{abstract}
We present a novel simulation-based hybrid emulator approach that maximally derives cosmological and Halo Occupation Distribution (HOD) information from non-linear galaxy clustering, with sufficient precision for DESI Year 1 (Y1) analysis. 
Our hybrid approach first samples the HOD space on a fixed cosmological simulation grid to constrain the high-likelihood region of cosmology+HOD parameter space, and then constructs the emulator within this constrained region. This approach significantly reduces the parameter volume emulated over, thus achieving much smaller emulator errors with fixed number of training points. We demonstrate that this combined with state-of-the-art simulations result in tight emulator errors comparable to expected DESI Y1 LRG sample variance. We leverage the new \textsc{AbacusSummit} simulations and apply our hybrid approach to CMASS non-linear galaxy clustering data. We infer constraints on $\sigma_8 = 0.762\pm0.024$ and $f\sigma_8(z_\mathrm{eff} = 0.52) = 0.444\pm0.016$, the tightest among contemporary galaxy clustering studies. We also demonstrate that our $f\sigma_8$ constraint is robust against secondary biases and other HOD model choices, a critical first step towards showcasing the robust cosmology information accessible in non-linear scales. We speculate that the additional statistical power of DESI Y1 should tighten the growth rate constraints by at least another 50-60$\%$, significantly elucidating any potential tension with Planck. \red{We also address the ``lensing is low'' tension, which we find to be in the same direction as a potential tension in $f\sigma_8$. We show that the combined effect of a lower $f\sigma_8$ and environment-based bias accounts for approximately $50\%$ of the discrepancy.} 
\end{abstract}




\begingroup
\let\clearpage\relax
\endgroup
\newpage

\section{Introduction}
The spatial distribution of galaxies presents one of the most powerful probes of the fundamental properties of the universe. Over the last few decades, galaxy clustering has emerged as an essential tool in constraining cosmology and galaxy evolution, especially as past and current cosmological surveys have increased the amount of galaxy clustering data available by orders of magnitude. Wide-area spectroscopic surveys such as the SDSS-III Baryon Oscillation Spectroscopic Survey \citep[BOSS;][]{2013Dawson}, the
SDSS-IV extended Baryon Oscillation Spectroscopic Survey \citep[eBOSS;][]{2016Dawson}, and WiggleZ \citep[][]{2011Blake} have proven to be particularly constraining, yielding groundbreaking results on the expansion history of the universe and the growth of structure. Upcoming spectroscopic surveys such as the Dark Energy Spectroscopic Instrument \citep[DESI;][]{2016DESI}, the Subaru Prime Focus Spectrograph \citep[PFS;][]{2014Takada}, the ESA \textit{Euclid} satellite mission \citep[][]{2011Laureijs}, and the NASA \textit{Roman Space Telescope} \citep[WMAP;][]{2013Spergel} should enable the next giant leap in both the volume and depth in our mapping of the universe. 

Galaxy clustering is a particularly sensitive tracer of the growth of cosmic structure, which has recently emerged as a potential tension point in the standard $\Lambda$CDM paradigm. Specifically, galaxy clustering studies directly constrain the growth rate parameter combination $f\sigma_8$, where $f$ is the linear growth rate of structure, and $\sigma_8$ quantifies the normalisation of the matter power spectrum. The tension in $f\sigma_8$ arises when comparing late-time ($z<1$) galaxy clustering and lensing constraints to the CMB constraints, where the late-time constraints are systematically lower than CMB with mild statistical significance. \citet{2013Macaulay} first reported a lower than expected growth rate by combining several early large-scale RSD measurements. A series of subsequent lensing studies claimed a tension of around $2\sigma$ \citep[e.g.][]{2015MacCrann, 2015Battye, 2020Hildebrandt, 2022Amon}, which are further corroborated by recent clustering constraints on growth rate that systematically bias low by 1--2$\sigma$ \citep[e.g.][]{2014Sanchez, 2016Gilmarin, 2017Nesseris, 2021deMattia}. We refer the readers to \citet{2022Abdalla} for a comprehensive review of this tension and potential solutions. 

To extract cosmological information from galaxy clustering, one popular approach is to isolate out the so-called ``standard ruler'' features, such as the baryon acoustic peak, from the full-shape clustering. These standard rulers are robust to most observational systematics \citep{2012Ross, 2015bRoss}, but capture only a limited amount of information contained in the full clustering measurement. Another approach is to compare the measured full-shape clustering directly to a theoretical template, which in principle captures the full information content of the galaxy clustering signal. This approach can place particularly stringent constraints on the growth of structure by jointly probing both the matter density field and the matter velocity field, as the full-shape galaxy clustering is modulated by velocity distortion effects such as the Kaiser effect \citep{1987Kaiser} and the \textit{finger-of-god} (FoG) effect, collectively known as redshift-space distortions (RSD). 

RSD studies have been mostly limited to very large scales, as the theory templates rely on variations of high-order perturbation theories, which are only reliable beyond approximately 30--50$h^{-1}$Mpc \citep[][]{2009Carlson, 2013Carlson}. In fact, even on very large scales, RSD measurements can be modulated by a variety of non-linear effects \citep{1994Cole, 2004Scoccimarro, 2006Tinker}. However, modern cosmological surveys are designed in a way that their galaxy clustering measurements are most accurate at scales of a few megaparsecs, far below the limits of perturbative models. Thus, to fully take advantage of the rich cosmological and astrophysical information contained in current and upcoming galaxy survey data, new models need to be developed that can accurately predict clustering on much smaller scales. 

Small-scale clustering is also much more sensitive to non-linear growth, and thus should be significantly more constraining on $f\sigma_8$. In fact, \citet{2019Zhai} showed that the constraining power on $f\sigma_8$ increases monotonically as smaller scales are included. With the same data set, the projected error bar on $f\sigma_8$ from a small-scale analysis is half of what is possible with a perturbative analysis on large scales. Thus, in principle, a robust small-scale analysis with DESI data can drastically increase the statistical significance of a potential tension in $f\sigma_8$. 

However, modeling structure on small scales is challenging. The small-scale RSD is complicated not only by high-order contributions of both the density and velocity fields, but also non-perturbative effects arising from the dynamics beyond shell crossing, i.e., formation and evolution of galaxies (or dark matter halos) and baryonic feedback. To tackle these complexities, a new class of methods relying on simulations instead of analytic theories have arisen \citep[e.g.][]{2019Zhai, 2021Lange, 2021Chapman, 2021Kobayashi}. The key advantage of simulation-based methods is that N-body simulations can precisely capture the non-linear evolution of dark matter density and velocity fields, given sufficient computational resources. However, N-body simulations only simulate the gravitational growth of the total matter field, so we still need a robust galaxy--dark matter connection model that ``paints'' galaxies on top of the simulated matter density field. An additional challenge with simulation-based methods is the fact that simulations of sufficiently large volume (larger than survey volume) and precision are expensive to run, and can only be run at fixed cosmology, resulting in a very sparse sampling (often $\sim 10$ per parameter direction) of cosmology parameter space. 

To bridge the galaxy--halo connection, several empirical models have emerged as being particularly effective at producing the correct small-scale galaxy clustering. \red{One such model is sub-halo abundance matching \citep[SHAM;][]{2006Conroy, 2013Reddick, 2016Chaves}, which in its most basic form, relies on the assumption that some galaxy property (such as luminosity or stellar mass) is monotonically related to some halo properties (such as virial mass or maximum circular velocity) with scatter. More complex versions of this modeling approach have also parameterized for example tidal disruption and assembly bias \citep[e.g.][]{2017Lehmann, 2021Contreras, 2022Hearin}.} \red{Another} model commonly used in modern cosmological simulations is the halo occupation distribution model \citep[HOD;][]{2002Berlind, 2005Zheng, 2007bZheng, 2016Hearin, 2018Yuan}, which ties galaxy occupation probabilistically to halo properties through a set of parametrized functions. We refer the readers to \citet{2018Wechsler} for a comprehensive review of this topic. 

The issue of sparse cosmology sampling can be addressed with interpolative surrogate models known as emulators, as demonstrated with various degrees of success in recent studies \citep[e.g. ][]{2009Heitmann, 2010Lawrence, 2014Heitmann, 2019Zhai, 2022Zhai, 2019Lange, 2021Lange, 2021Kobayashi}. These emulator models utilize flexible parametrized models such as polynomial expansions and Gaussian processes to interpolate between the model predictions at the available cosmologies to enable dense sampling of the cosmology likelihood space, but at the cost of additional errors due to uncertainties in the model. Thus, the key challenge in emulator studies is to measure and minimize such emulator errors. 

While recent emulator studies have demonstrated sufficient precision to analyze existing datasets such as BOSS CMASS and LOWZ, they are still held back by the sparse cosmology+HOD sampling and limited simulation volume. As a result, the emulator errors are too large to take advantage of the effective volumes of upcoming surveys such as DESI. Another issue with existing emulator analyses is the reliance on a single HOD parametrization. There is significant uncertainty associated with what the correct HOD form is, and a key test of the cosmological information content of small-scale galaxy clustering is to show that the cosmology constraints are independent of the assumed HOD model. 

In this paper, we develop a new hybrid emulator framework that aims to tackle these issues by combining emulation with direct Markov chain Monte Carlo (MCMC) sampling, and by taking advantage of new state-of-the-art simulations and highly efficient HOD codes. To demonstrate the efficacy of our new method, we apply it to existing CMASS data and achieve stringent joint constraints on cosmology and HOD. 

Specifically, to minimize emulator errors, we adopt the largest N-body simulation suite available, \textsc{AbacusSummit}, which is designed to satisfy and exceed the simulation requirement of DESI \citep[][]{2021Maksimova}. These simulations offer more than an order of magnitude more in total volume than the previous state-of-the-art emulator simulations \citep[e.g.][]{2019DeRose}, significantly suppressing sample variance in our emulators. To further reduce emulator errors, we adopt a hybrid approach,  where we first run MCMC chains to identify high-likelihood regions of parameter space and only emulate the model prediction in the identified regions. This is in contrast to existing analyses where the training samples were selected blind of the data and are thus constructed to cover expansive parameter space. By training our emulator in a much smaller data-constrained regions of cosmology+HOD parameter space, we increase the density of training points and thus achieve smaller interpolation errors. We show that our hybrid approach should be sufficiently precise to utilize the effective volume of the DESI Luminous Red Galaxy (LRG) sample. 

Our approach also enables testing different HOD models by using the \ahod\ package \citep[][]{2021bYuan}, which contains a large set of physically extensions to the standard HOD model that can be switched on and off without penalizing the computational efficiency. For the bulk of this paper, we carry our analysis through with two different HOD models, specifically one with and one without environment-based secondary bias. We also test our key cosmological constraints against additional HOD model extensions and show that our cosmology results are robust against such variations. Finally, a key advantage of \ahod\ lies in its high efficiency, which both enables the MCMC step and increases the size of the HOD training set we can generate. 


This paper is structured as follows. Section~\ref{sec:theory} describes the simulation-based model template we construct to compare with observed galaxy clustering measurements. In section~\ref{sec:data}, we present the target galaxy clustering measurement we analyze for this paper. In section~\ref{sec:emulator}, we describe in detail the hybrid MCMC+emulator methodology and the corresponding inference pipeline. Section~\ref{sec:results} presents the main science results of this paper, summarizing the cosmology+HOD posterior constraints. In section~\ref{sec:discussion}, we compare our results to relevant studies and further evaluate our results in a broader context. Finally, we conclude in section~\ref{sec:conclude}.

\section{Simulation-based model template}
\label{sec:theory}
In this section, we describe our simulation-based model template for 3D galaxy clustering measurements on small scales. We first introduce the \textsc{AbacusSummit} simulations, which generates the dark matter density and velocity fields at different cosmologies. We then introduce the \ahod\ framework that populates the dark matter fields with galaxies and computes the predicted galaxy clustering. 
\subsection{The \textsc{AbacusSummit} simulations}
\label{sec:simulation}

The \textsc{AbacusSummit} simulation suite \citep[][]{2021Maksimova} is a set of large, high-accuracy cosmological N-body simulations using the \textsc{Abacus} N-body code \citep{2019Garrison, 2021bGarrison}, designed to meet and exceed the Cosmological Simulation Requirements of the Dark Energy Spectroscopic Instrument (DESI) survey \citep{2013arXiv1308.0847L}. \textsc{AbacusSummit} consists of over 150 simulations, containing approximately 60 trillion particles at 97 different cosmologies. 
For this analysis, we use exclusively the ``base'' configuration boxes within the simulation suite, each of which contains $6912^3$ particles within a $(2h^{-1}$Gpc$)^3$ volume, corresponding to a particle mass of $2.1 \times 10^9 h^{-1}M_\odot$. \footnote{For more details, see \url{https://abacussummit.readthedocs.io/en/latest/abacussummit.html}}

The \textsc{AbacusSummit} suite also uses a new specialized spherical-overdensity based halo finder known as {\sc CompaSO} \citep{2021Hadzhiyska}. Specifically, {\sc CompaSO} is a highly efficient on-the-fly group finder specifically designed
for the \textsc{AbacusSummit} simulations. 
{\sc CompaSO} builds on the existing 
spherical overdensity (SO) algorithm
by taking into consideration the tidal radius
around a smaller halo before competitively
assigning halo membership to the particles
in an effort to more effectively deblend halos.
Among other features, the {\sc CompaSO} finder also
allows for the formation of new halos on the 
outskirts of growing halos, which alleviates
a known issue of configuration-space halo 
finders of failing to identify halos close to
the \red{center of mass} of larger halos. 

The {\sc CompaSO} halos also leverage a post-processing ``cleaning'' procedure that utilizes the halo merger trees to ``re-merge'' a subset of halos \citep[][]{2021Bose}. This is done both to remove over-deblended halos in the spherical overdensity finder, and to intentionally merge physically associated halos that have merged and then physically separated. An example of such dissociation is what is known as splashback \citep[e.g.][]{2014Diemer, 2015bMore, 2016More}, where halos that were once part of a larger halos have since exited following at least one orbital passage within their former hosts. In \citet{2021Bose}, the authors found that re-merging such halos significantly improves the fidelity of the halo catalog, and the resulting ``cleaned'' halo catalog achieves significantly better fits on data in an HOD analysis. The fits presented this paper are carried out with the cleaned halo catalogs. 

\subsubsection{The cosmology grid}
For this analysis, we mainly utilize two sets of base boxes. The first set is a set of 85 base boxes run at 85 distinct cosmologies, at fixed initial phase. The second set is a set of 25 base boxes run at the same fiducial cosmology \citep[P18;][also known as c000 later on]{2020Planck}, but at 25 different initial phases. The first set is essential for building up the training set for the cosmology+HOD emulator (see section~\ref{subsec:method_emulator}). The second set is important for quantifying the sample variance and potential biases (see section~\ref{subsec:cov_bias}). The 85 cosmologies are tagged c000-181 non-consecutively while 25 phases are tagged ph000-024. 

We now briefly describe the 85 different cosmologies and how they are chosen (see \citealt{2021Maksimova} for details). The cosmology parameter basis used for our emulator includes 8 parameters: the baryon density $\omega_b = \Omega_b h^2$, the cold dark matter density $\omega_\mathrm{cdm} = \Omega_\mathrm{cdm}h^2$, the amplitude of structure $\sigma_8$, the spectral tilt $n_s$, running of the spectral tilt $\alpha_s$, the density of massless relics $N_\mathrm{eff}$, and dark energy equation of state parameters $w_0$ and $w_a$ ($w(a) = w_0+(1-a)w_a$). 

A set of observational constraints where followed in the design of the parameter grid. For example, $w_a$ is varied while holding the equation of state
at $z = 0.333$ constant, so that the low redshift cosmic distance scale remains less changed. Similarly, changes in $N_\mathrm{eff}$ are compensated by changes in $\omega_\mathrm{cdm}$ and $n_s$, so as to leave the CMB less changed. A flat spatial curvature is assumed for all cosmologies, and the Hubble constant $H_0$ is chosen to match the CMB acoustic scale $\theta_*$ to P18 measurement. 

\begin{figure*}
    \centering
    \hspace*{-0.6cm}
    \includegraphics[width = 7.2in]{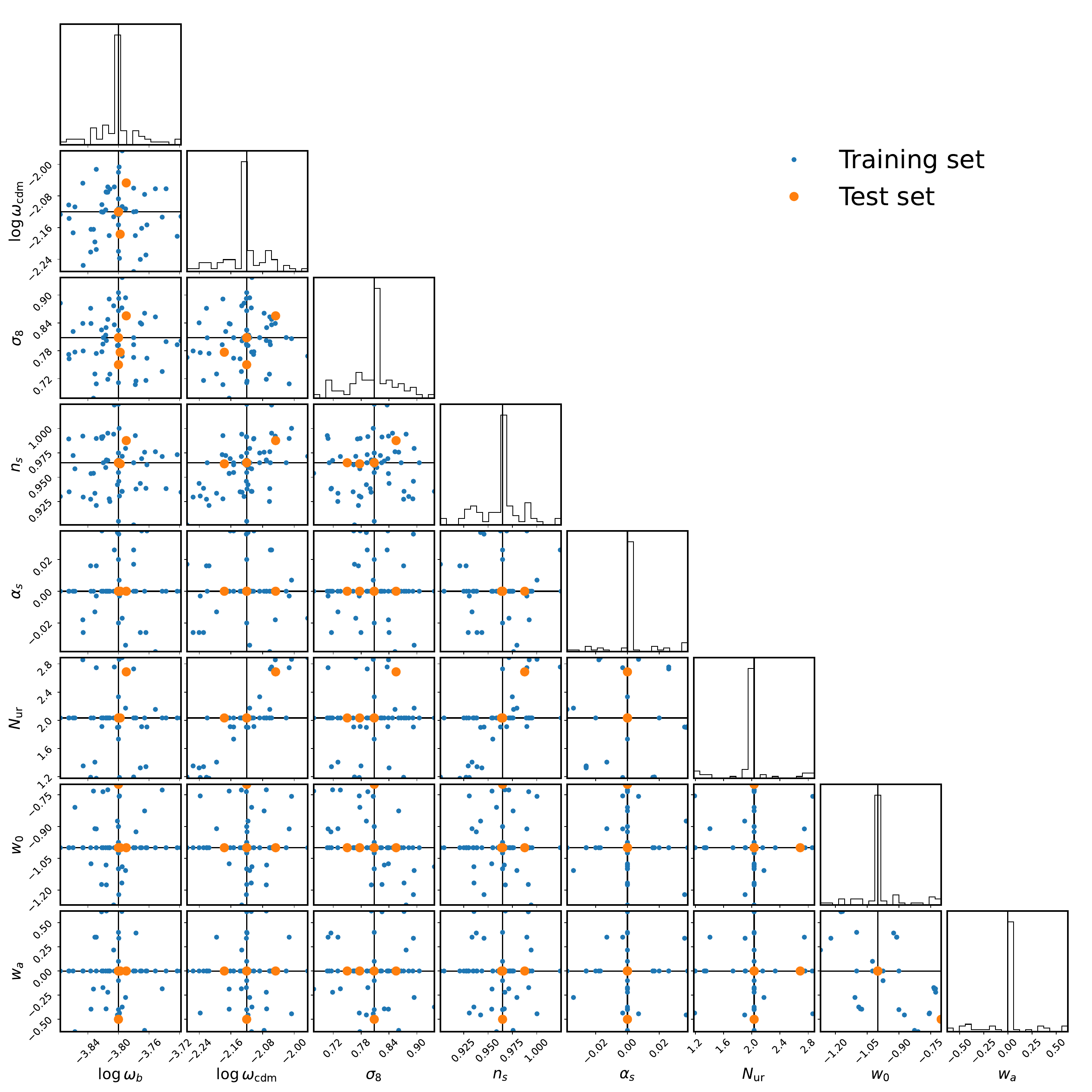}
    \vspace{-0.3cm}
    \caption{The cosmology grid used to train/test our emulators. There are a total of 85 different cosmologies spanning an 8-dimensional parameter space. The Hubble constant $H_0$ is chosen to match the CMB acoustic scale $\theta_*$ to P18 measurement. The black lines denote the central Planck 2018 cosmology. \red{5 cosmologies (c000-c004), highlighted in orange, are reserved for leave-one-out tests (refer to section~\ref{subsec:emu_test}).}}
    \label{fig:cosmo_grid}
\end{figure*}

The different cosmologies are indexed by \textsc{cXXX}, where \textsc{XXX} ranges from 000 to 181. The details of each cosmology are described on the \textsc{AbacusSummit} website \footnote{\url{https://abacussummit.readthedocs.io/en/latest/cosmologies.html}}. Figure~\ref{fig:cosmo_grid} visualizes the 85 different cosmologies relative to P18 shown in black lines. \red{Note that 5 cosmologies (c000-c004) are highlighted as the test set. Their usage is described in section~\ref{subsec:emu_test}.}
We briefly summarize the cosmology choices here. 

\textsc{c000} corresponds to the fiducial P18 cosmology, specifically the mean estimates of the Planck TT,TE,EE+lowE+lensing likelihood chains. The 25 boxes with different phases are at \textsc{c000}.

\textsc{c001-004} correspond to four secondary cosmologies. Specifically, a WMAP9 + ACT + SPT \citep[][]{2017Calabrese}, a thawing dark energy model ($w_0 = -0.7, w_a = -0.5$), a model with extra relativistic density ($N_\mathrm{eff} = 3.7$), and a model with lower amplitude of clustering (c000 but with $\sigma_8$ = 0.75). 

\textsc{c100-126} is a linear derivative grid set up around \textsc{c000}, with symmetric pairs along all 8 parameter axes. The grid also includes additional pairs along $\omega_\mathrm{cdm}$, $\sigma_8$, $n_s$, $w_0$, and $w_a$ with smaller step sizes. 

\textsc{c130-181} forms an emulator grid that provides a wider coverage of the 8-dimensional parameter space. This was done by placing electrostatic points on the surface of an 8-dimensional ellipsoid, whose extent was chosen to be 3 to 8 standard deviations beyond current constraints from the combination of CMB and large-scale structure data. The grid excludes anti-podal reflected points and includes extra excursion along $\sigma_8$.

\subsection{galaxy--halo connection modeling}
\label{sec:hod}
To propagate the simulated matter density field to galaxy distributions, we adopt the Halo Occupation Distribution model (HOD), which probabilistically populate dark matter halos with galaxies according to a set of halo properties. Statistically, the HOD can be summarized as a probabilitistic distribution $P(n_g|\boldsymbol{X}_h)$, where $n_g$ is the number of galaxies of the given halo, and $\boldsymbol{X}_h$ is some set of halo properties.

In the standard or vanilla HOD model, halo mass is assumed to be the only relevant halo property $\boldsymbol{X}_h = {M_h}$ \citep{2005Zheng, 2007bZheng}. This standard HOD separates the galaxies into central and satellite galaxies, and assumes the central galaxy occupation to follow a Bernoulli distribution where the satellites follow a Poisson distribution. For LRGs, the standard HOD assumes a 5-parameter model for the mean expected number of central and satellite galaxies per halo given halo mass. \red{In this analysis, we follow the same functional form as \citet{2005Zheng, 2007bZheng} but adopt the specific parameterization of \citet{2015Kwan}}:
\begin{align}
    \bar{n}_{\mathrm{cent}}^{\mathrm{LRG}}(M) & = \frac{1}{2}\mathrm{erfc} \left[\frac{\log_{10}(M_{\mathrm{cut}}/M)}{\sqrt{2}\sigma}\right], \label{equ:zheng_hod_cent}\\
    \bar{n}_{\mathrm{sat}}^{\mathrm{LRG}}(M) & = \left[\frac{M-\kappa M_{\mathrm{cut}}}{M_1}\right]^{\alpha}\bar{n}_{\mathrm{cent}}^{\mathrm{LRG}}(M),
    \label{equ:zheng_hod_sat}
\end{align}
where the five parameters characterizing the model are $M_{\mathrm{cut}}, M_1, \sigma, \alpha, \kappa$. $M_{\mathrm{cut}}$ characterizes the minimum halo mass to host a central galaxy. $M_1$ characterizes the typical halo mass that hosts one satellite galaxy. $\sigma$ describes the steepness of the transition from 0 to 1 in the number of central galaxies. $\alpha$ is the power law index on the number of satellite galaxies. $\kappa M_\mathrm{cut}$ gives the minimum halo mass to host a satellite galaxy.
We have added a modulation term $\bar{n}_{\mathrm{cent}}^{\mathrm{LRG}}(M)$ to the satellite occupation function to remove satellites from halos without centrals. 

For this analysis, we utilize the \ahod\ framework, a highly efficient HOD implementation that enables a large set of HOD extensions \citep[][]{2021bYuan}. The code is publicly available as a part of the \textsc{abacusutils} package at \url{http://https://github.com/abacusorg/abacusutils}. Example usage can be found at \url{https://abacusutils.readthedocs.io/en/latest/hod.html}. In the following two sub-sub-sections, we describe its efficiency and some relevant physically-motivated model extensions.

\subsubsection{\ahod\ Efficiency}

While the large volume of \textsc{AbacusSummit} is excellent for suppressing sample variance, but it also makes HOD evaluations painfully slow. Thus, a fast HOD implementation is essential for sampling a high-dimensional extended HOD + cosmology parameter spaces. \ahod\ achieves high efficiency through several key optimizations. First of all, it does a compression step by downsampling the halos and particles, but doing so in a halo mass dependent fashion in order to avoid introducing extra shot noise. Secondly, the code is accelerated with \textsc{numba} and highly parallelized to take advantage of all available processor cores. Finally, the code minimizes I/O by loading all downsampled halos and particles in memory during initialization and subsequently conducts all array manipulations in-place in memory. \ahod\ also utilizes the highly-optimized grid-based \textsc{Corrfunc} code \citep{2020Sinha} for fast correlation function calculations. For CMASS LRG density, a full HOD evaluation including the calculation of the correlation function on a base box takes approximately 0.2 seconds on a 32 core desktop machine. The performance of \ahod\ is described in more details in section~4.4 of \citet{2021bYuan}. 

\subsubsection{HOD extensions}
\label{subsubsec:hod_Ext}

\ahod\ also includes a rich set of HOD extensions beyond the standard 5-parameter HOD model \citep[$\log_{10}M_\mathrm{cut}$, $\log_{10}M_1$, $\sigma$, $\alpha$, $\kappa$, see][]{2007bZheng}. We summarize the relevant extensions here:
\begin{itemize}

    \item \texttt{$\alpha_\mathrm{vel, c}$} is the central velocity bias parameter, which modulates the peculiar velocity of the central galaxy relative to the halo center \red{(center of mass of the central subhalo)}. $\alpha_\mathrm{vel, c} = 0$ indicates no central velocity bias, i.e. centrals perfectly track the velocity of halo centers. 
    \item \texttt{$\alpha_\mathrm{vel, s}$} is the satellite velocity bias parameter, which modulates how the satellite galaxy peculiar velocity deviates from that of the local dark matter particle. $\alpha_\mathrm{vel, s} = 1$ indicates no satellite velocity bias, i.e. satellites perfectly track the velocity of their underlying particles.  
    \item \texttt{$s$} is the satellite profile bias parameter, which modulates how the radial distribution of satellite galaxies within haloes deviate from the radial profile of the halo. $s = 0$ indicates no radial bias, i.e. satellites are uniformly assigned to halo particles. $s > 0$ indicates a more extended (less concentrated) profile of satellites relative to the halo, and vice versa. 
    \item \texttt{$A_\mathrm{cent}$} or \texttt{$A_\mathrm{sat}$} are the concentration-based secondary bias parameters for centrals and satellites, respectively. Also known as galaxy assembly bias parameters. $A_\mathrm{cent} = 0, A_\mathrm{sat} = 0$ indicate no concentration-based secondary bias in the centrals and satellites occupation, respectively. A positive $A$ indicates a preference for lower concentration halos, and vice versa. 
    \item \texttt{$B_\mathrm{cent}$} or \texttt{$B_\mathrm{sat}$} are the environment-based secondary bias parameters for centrals and satellites, respectively. The environment is defined as the mass density within a $r_\mathrm{env} = 5h^{-1}$Mpc tophat of the halo center, excluding the halo itself. $B_\mathrm{cent} = 0, B_\mathrm{sat} = 0$ indicate no environment-based secondary bias. A positive $B$ indicates a preference for halos in less dense environments, and vice versa. 
\end{itemize}

Finally, the \ahod\ framework also offers multi-tracer capabilities, i.e. ability to simultaneously model the occupation of multiple galaxy types. While multi-tracer is an essential component of \ahod, we do not describe it further since it is not relevant for this paper. 

For this paper, our fiducial analysis invokes two sets of the extensions described above: central and satellite velocity bias ($\alpha_\mathrm{vel, c}$, $\alpha_\mathrm{vel, s}$), and central and satellite environment-based secondary bias ($B_\mathrm{cent}$, $B_\mathrm{sat}$). We focus on these extensions as they are very well motivated by observations and simulations. Velocity bias has been shown to be a necessary ingredient in modeling BOSS galaxy redshift-space clustering on small scales \citep[e.g.][]{2015aGuo, 2021bYuan}, where various studies have found consensus values of $\alpha_\mathrm{vel, c}\approx 0.2$ and $\alpha_\mathrm{vel, s}\approx 1$. Velocity bias has also been identified in hydrodynamical simulations and measured to be consistent with observational constraints \citep[e.g.][]{2022Yuan, 2017Ye}. 

In terms of secondary bias, traditionally, galaxy concentration has been used as the sole marker of galaxy secondary bias (also known as galaxy assembly bias) due to the concentration's correlation with halo age \citep[e.g.][]{2002Wechsler, 2007Croton, 2007Gao}, with older halos having higher concentrations and vice versa.
However, a series of recent studies found local environment of the halo to be an excellent tracer of galaxy secondary bias based on hydrodynamical simulations and semi-analytic models \citep[e.g.][]{2022Yuan, 2020Hadzhiyska, 2020Xu, 2021Xu, 2021Delgado}. Specifically, \citet{2020Hadzhiyska} and \citet{2020Xu} both systematically tested the effectiveness of various secondary HOD dependencies in capturing the secondary bias signature, \citet{2020Hadzhiyska} through the \textsc{IllustrisTNG} hydrodynamical simulations \citep[e.g.][]{2018Pillepich,2018Springel,2018Nelson} and \citet{2020Xu} through semi-analytical models. Both studies found the halo environment to be the best at reproducing the excess clustering due to secondary biases. More recently, \citet{2021Xu} and \citet{2021Delgado} used random forests to systematically identify the most important halo properties in an HOD framework, using hydrodynamical simulations and semi-analytic models, respectively. Both studies again found halo mass and halo environment to be by far the two most important galaxy occupation dependencies. Motivated by these studies, we opt to include the environment-based secondary bias parameters ($B_\mathrm{cent}$, $B_\mathrm{sat}$) in our fiducial HOD model. However, we do test the concentration-based secondary bias and other HOD extensions in section~\ref{subsec:additional_extensions}.

\bigbreak

Finally, having generated the galaxy catalog given a simulation volume, we utilize the fast grid-based \textsc{Corrfunc} code \citep{2020Sinha} to compute the full-shape galaxy clustering. For this analysis, we consider specifically the redshift-space 2PCF $\xi(r_p, \pi)$, which decomposes the 3D 2-point correlation function into the basis of perpendicular separation $r_p$ vs. the line-of-sight (LoS) separation $\pi$. This is different than the more commonly used multipole decomposition. In principle, the two decompositions contain the same information, but \red{we conjecture that $(r_p, \pi)$ forms the more convenient basis in separating out the small-scale FoG effect and systematic effects such as fiber-collision correction issues. We reserve the detailed comparison between the two bases for a separate study.}

\section{BOSS CMASS Data}
\label{sec:data}
While the hybrid emulator framework we propose in this paper is designed for upcoming surveys such as DESI, we utilize the existing BOSS CMASS dataset to demonstrate its constraining power. 

The Baryon Oscillation Spectroscopic Survey \citep[BOSS; ][]{2012Bolton, 2013Dawson} is part of the SDSS-III programme \citep{2011Eisenstein}. BOSS Data Release 12 (DR12) provides redshifts for 1.5 million galaxies in an effective area of 9329 square degrees divided into two samples: LOWZ and CMASS. The CMASS sample is designed to isolate galaxies of approximately constant mass at higher redshift ($z > 0.4$), most of them being also Luminous Red Galaxies \citep[LRGs,][]{2016Reid, 2016Torres}. The survey footprint is divided into chunks which are covered in overlapping plates of radius $\sim$1.49 degrees. Each plate can house up to 1000 fibres, but due to the finite size of the fibre housing, no two fibres can be placed closer than $62$ arcsec, referred to as the fibre collision scale \citep{2012Guo}. 

For this analysis, we limit our measurements to the CMASS sample between redshift $0.46 < z < 0.6$ in DR12. We choose these moderate redshift ranges for completeness and to minimize the systematics due to redshift evolution. Applying this redshift range to both the North and South Galactic caps gives a total of approximately 600,000 galaxies in our CMASS sample. We then convert angular positions and redshift to 3D comvoing positions via a reference cosmology set to P18 values. In principle, we would need to account for a potential mismatch between the assumed cosmology and the inferred cosmology, an effect known as the Alcock--Paczynski (AP) effect \citep[][]{1979Alcock}. However, recent small-scale RSD studies have found this effect to have negligible impact on the final parameter constraints \citep[][Zhai et al. in prep]{2021Chapman}. Thus, we do not explicitly model this effect for this analysis. The resulting sample has an average galaxy number density of $n_\mathrm{CMASS} = (3.01\pm 0.03)\times 10^{-4} h^{3}$Mpc$^{-3}$. 

\begin{figure*}
    \centering
    \hspace*{-0.6cm}
    \includegraphics[width = 6.5in]{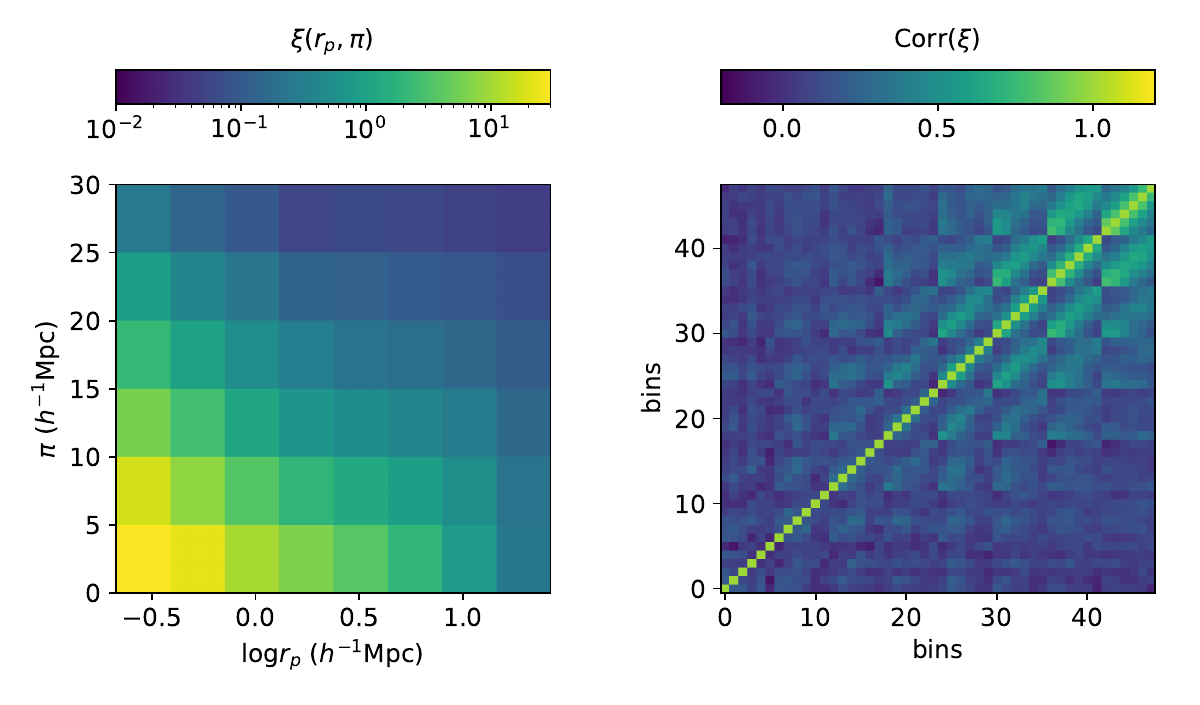}
    \vspace{-0.3cm}
    \caption{The redshift-space 2-point correlation function (left) of the BOSS CMASS DR12 galaxies at $0.46 < z < 0.6$ and its corresponding covariance matrix (right). $r_p$ is the transverse comoving distance between galaxies. $\pi$ is the LoS comoving distance between galaxies. The right hand side shows the correlation matrix,  calculated from 400 jackknife sub-samples. The bins are formed by flattening the $\xi$ bins column-by-column, with large bin number corresponding to large $r_p$. For example, the bin 0-5 correspond to the first $r_p$ bin but increasing $\pi$.}
    \label{fig:xi_boss}
\end{figure*}

To compute the redshift-space 2PCF $\xi(r_p, \pi)$ on CMASS, we use the \citet{1993Landy} estimator:
\begin{equation}
    \xi(r_p, \pi) = \frac{DD - 2DR + RR}{RR},
    \label{equ:xi_def}
\end{equation}
where $DD$, $DR$, and $RR$ are the normalized numbers of data-data, data-random, and random-random pair counts in each bin of $(r_p, \pi)$, and $r_p$ and $\pi$ are transverse and LoS separations in comoving units. For this paper, we choose a coarse binning to ensure reasonable accuracy on the covariance matrix, with 8 logarithmically-spaced bins between 0.169$h^{-1}$Mpc and 30$h^{-1}$Mpc in the transverse direction, and 6 linearly-spaced bins between 0 and 30$h^{-1}$Mpc bins along the LoS direction. The measurement on our CMASS sample is visualized in the left panel of Figure~\ref{fig:xi_boss}.

We have corrected the fibre collision effect following the method of \cite{2012Guo}, by separating galaxies into collided and de-collided populations and assuming those collided galaxies with measured redshifts in the plate-overlap regions are representative of the overall collided population. The final corrected correlation function can be obtained by summing up the contributions from the two populations.

The covariance matrix is calculated from 400 jackknife samples. The number of jackknife is chosen to ensure that the covariance matrix is well-conditioned while also leaving enough volume in each jackknife. The correlation matrix shown in the right panel of  Figure~\ref{fig:xi_boss} is defined relative to the covariance matrix as 
\begin{equation}
\mathrm{Corr}(\xi)_{ij} = \mathrm{Cov}(\xi)_{ij}/\sqrt{\mathrm{Cov}(\xi)_{ii}\mathrm{Cov}(\xi)_{jj}}.
\label{equ:corr}
\end{equation}

The $x$ and $y$ axes show the same bins as on the left panel, flattened in a column-by-column fashion such that the transverse separation $r_p$ increases with bin number. Overall, we see that the off-diagonal power is relatively small at small transverse scales and becomes more significant at larger transverse scales. This suggests that the error at small $r_p$ is shot-noise dominated, while sample variance becomes dominant at large $r_p$. We continue the discussion of the covariance matrices in section~\ref{subsec:cov_bias}. 

\section{Likelihood modeling}
\label{sec:emulator}
While N-body simulations such as \textsc{AbacusSummit} provide high-fidelity clustering predictions down to the one-halo scale, simulations are expensive to produce and only produce a sparse sampling of the cosmology parameter space. However, in order to constrain cosmology, we need a robust forward model that can predict clustering given arbitrary reasonable cosmology and HOD parameter combination within a designed range. This is one of the fundamental challenges of simulation-based modeling of cosmological observables. 

\subsection{Existing methods}

There have been two well-developed approaches applied to overcome the issue of sparse sampling of cosmology. The more popular approach is through the use of approximate surrogate models, or emulators, to interpolate between the predictions of the sparse samples. Specifically, in this approach one assumes that the model prediction is smooth as a function of cosmological parameters, and thus trains a regularized analytical model (the emulator) on the cosmology and HOD samples. This approach has started to mature over the last decade through a series of progressively more sophisticated studies. \citet{2009Heitmann} and \citet{2014Heitmann} represent one of the earliest simulation-based emulator works, where the authors emulated the matter power spectrum with Gaussian Processes (GP) based surrogate models, achieving approximately $1\%$ error. More recent works have focused on emulation of galaxy clustering statistics, for which one has to emulate the galaxy--halo connection model (mostly commonly an HOD) in addition to cosmology. In \citet{2019Wibking} and \citet{2019bWibking}, the authors used a GP model to emulate the projected clustering and galaxy--galaxy lensing as a function of cosmology and HOD, achieving an out-sample error of $\sim 3\%$ ($68$ C.L.) on the projected correlation function.
The \textsc{Aemulus} project \citep[e.g.][Zhai et al. in prep.]{2019Zhai, 2019DeRose} use a similar GP-based emulator but with a specialized set of N-body simulations specifically designed for emulation. Their analysis achieved percent-level emulator error on projected galaxy clustering and redshift-space multipoles within $r < 10 h^{-1}$Mpc, but looses accuracy at larger scales. 
The Dark Emulator project \citep{2019Nishimichi, 2021Miyatake} adopts a slightly different approach where they emulator halo clustering statistics, which is then convolved with an analytic HOD model for galaxy clustering prediction. This approach does simplify the emulation to a certain extent, but limits the amount of flexibility in the HOD model. 

The second approach seeks to directly interpolate the marginalized likelihood as a function of cosmology. Specifically, for galaxy clustering inference, one would calculate the likelihood marginalized over all HOD realizations at each cosmology, and then interpolate the marginalized likelihood as a function of cosmological parameters. This approach was first developed in \citet{2019Lange}, known there as cosmological evidence modeling, and then successfully applied to BOSS LOWZ galaxy clustering to obtain cosmology constraints \citep[][]{2021Lange}. This evidence modeling approach significantly simplifies the interpolation problem by removing the need to explicitly model HOD dependencies, allowing for arbitrarily complex HOD models. However, the downside of this approach is that it only allows for marginalized constraints on cosmology or HOD, but does not allow for joint cosmology+HOD parameter constraints. 

In this section, we describe in detail our new likelihood model and cosmology+HOD inference pipeline that improve upon these two existing approaches. 

\subsection{Hybrid MCMC + Emulator}
\label{subsec:method_emulator}

Our hybrid approach combines direct likelihood sampling through Markov chain Monte Carlo (MCMC) and an emulator interpolation scheme. Specifically, knowing the observed galaxy clustering data vector, we first fully sample the likelihood surface as a function of HOD parameters at each simulated cosmology with the publicly available \textsc{emcee} code \citep[][]{2013Foreman}. This is to identify the high-likelihood regions of parameter space that produce predictions similar to the observations, i.e. the typical set. Then we train our emulator only on these regions of parameter space to reduce the parameter volume we need to emulate over, thus increasing the density of training points and making the emulation problem easier. We see this as an methodological improvement compared to the standard emulator approach, where the emulator is trained over an expansive grid of HOD parameters at each cosmology, covering large regions of parameter space that is not supported by observations. 
Once we have identified the typical set, we can then uniformly sample our emulator training set from the MCMC samples at each cosmology. In spirit, this approach is similar to evidence modeling, except we interpolate the typical set at each cosmology instead of the marginalized likelihood. 

For this analysis, we adopt a Gaussian Processes (GP) based emulator model. For a more complete description of Gaussian Processes, we refer the reader to  \cite{williams2006gaussian}. Section~2.4 of \citet{2019Zhai} offers an abbreviated description of the core ideas of GPs in the context of cosmological emulators. For all Gaussian Process calculations in this work we use the \texttt{GPy} Python package \citep{gpy}. However, as demonstrated in previous emulator works, the standard GP models suffers from rather poor performance scaling as a function the number of training samples. Specifically, GP evaluations involve forming and inverting an $n \times n$ matrix where $n$ is the number of training points. This operation naively takes $O(n^2)$ memory and $O(n^3)$ time. Given current constraints on computing hardware (for this analysis, we are limited to 32-core nodes with 128GB memory), this limits a typical galaxy clustering emulator to use only a few thousand cosmology+HOD training points. 

Fortunately, we can outperform the naive scaling by exploiting structure in our training data distribution, in what is known as the Kronecker GP model. 
Specifically, if the space of HOD models in our training data is $X_\mathrm{HOD}$ and similarly for cosmological models $X_{\Omega}$, the full training parameter space $X$ can be written as a tensor product of the two spaces: 
\begin{equation}
    X = X_\mathrm{HOD} \otimes X_{\Omega}.
\end{equation}
As a result, a covariance kernel in $K \in X$ can be computed as a Kronecker product of $K_\mathrm{HOD} \in X_\mathrm{HOD}$ and $K_{\Omega} \in X_{\Omega}$
\begin{equation}
    K = K_\mathrm{HOD} \otimes K_{\Omega}.
\end{equation}
This structure, along with the following property of inverse Kronecker products
\begin{equation}
    \left( A \otimes B\right) ^ {-1} = A^{-1} \otimes B^{-1},
\end{equation}
enables the construction of a large GP as a product of smaller, more manageable GPs. We leverage this technique to make our emulators significantly more memory and time efficient. Specifically, we can improve our time complexity from $O(\Pi \; n_i^3)$ to $O(\Sigma n_i^3)$, where $n_i$ is the the number of training points in each parameter space. The details of this Kronecker GP regression are shown in \citep{stegle2011efficient}, and implemented in \texttt{GPy}'s \texttt{GPKroneckerGaussianRegression} class.

However, the separation of cosmology kernels and HOD kernels also means that Kronecker GP models necessarily require separate HOD training sets and cosmology training sets. In practice, it means that we need to train the GP model at the same set of HODs at each cosmology, while our MC sampling does not return the same HOD chains at different cosmologies. To overcome this issue, we uniformly sample 30 HODs from each MC chain and concatenate these subsamples across all 85 different cosmologies to arrive at a set of $30\times 85 = 2550$ HOD training points. Then we re-run these 2550 HODs at each of the 85 cosmology to form the final \red{2550 HOD $\times$ 85 cosmology} training set for the GP model. Effectively, instead of training our model over an arbitrary-shaped typical set in cosmology+HOD parameter space, we train our model over the minimum bounding box that envelops the typical set. While this somewhat compromises the spirit of finding the typical set in the first place, the benefit of being able to include an order-of-magnitude more HOD training points with Kronecker GPS outweighs the loss due to a somewhat enlarged training range. 
We also note that each HOD is also run 16 times with different random seeds and then averaged to suppress shot noise. 

To emulate the galaxy clustering function once we have compiled the training set, we have to choose a set of kernels for both the HOD covariance matrix and the cosmology covariance matrix. For this analysis, we decided on using the following kernel for both covariance matrices
\begin{equation}
\label{eqn:final_kernel}
k(x,x') =  k_\mathrm{exp}(x,x') + k_{3/2}(x,x') + k_\mathrm{white}(x,x').
\end{equation}
$x$ and $x'$ refer to two different sets of parameters. $k_\mathrm{exp}(x,x')$ is the commonly used squared exponential kernel. $k_{3/2}(x,x')$ is the Matern class kernel. $k_\mathrm{white}(x,x')$ is what is known as a white kernel, which accounts for uncertainties on the input parameters. We have also tested other kernel combinations, including ones used in existing cosmological emulators, and we found our kernel combination to yield the smallest out-sample errors given our training sets. 

We build a separate Kronecker GP for each ($r_p, \pi$) bin in our data vector (see Figure~\ref{fig:xi_boss}), \red{i.e. our final emulator model consists of $N_\mathrm{bin}$ number of independent emulators}. To optimize the values of our hyperparameters, we maximize the likelihood of each of our GP models with respect to our training data. We perform this operation using \texttt{GPy}'s \texttt{optimize\_restarts} routine, executing the optimization 5 times from random initialization and selecting the set with the highest likelihood. We also test different optimizer choices and found the TNC optimizer to be the most consistent in converging within a fixed time allocation. 

To summarize, our fiducial analysis uses 2550 HOD parameter combinations at 85 different cosmologies as the emulator training set. The emulator parameter basis include the 8 cosmological parameters varied in the \textsc{AbacusSummit} cosmology grid ($\omega_b$, $\omega_\mathrm{cdm}$, $\sigma_8$, $n_s$, $\alpha_s$, $N_\mathrm{eff}$, $w_0$, and $w_a$), and 9 HOD parameters ($\log_{10}M_\mathrm{cut}$, $\log_{10}M_1$, $\sigma$, $\alpha$, $\kappa$, $\alpha_\mathrm{vel, c}$, $\alpha_\mathrm{vel, s}$, $B_\mathrm{cent}$, $B_\mathrm{sat}$). We summarize the full parameter basis in Table~\ref{tab:parameters}. The emulator adopts the Kronecker GP model and is independently trained for each data vector bin.

\begin{table*}
\centering
\begin{tabular}{llcr}
\hline
& Parameter    & Meaning & Range \\
\hline
Cosmology & $\omega_{b}$      & The baryon energy density    & [0.0207,  0.0243]   \\
  & $\omega_\mathrm{cdm}$         & The cold dark matter energy density      & [0.103,  0.140]      \\
  &  $\sigma_8$ & The amplitude of the linear power spectrum evaluated at 8 Mpc $h^{-1}$    & [0.678,  0.938]     \\
   &  $n_{s}$       & The spectral index of the primordial power spectrum     & [0.901,  1.025]    \\
   &  $\alpha_{s}$       & The running of the spectral index    & [-0.038,  0.038]    \\
  &  $N_{\text{ur}}$ & The effective number of relativistic species  &  [1.177,  2.889] \\
   &  $w_0$ & The dark energy equation of state at $z=0$     & [-1.27, -0.70]   \\
   &  $w_a$ & The dark energy equation of state $a$ dependence    & [-0.628,  0.621]   \\

\hline 
  HOD &  $\log_{10}{M_\mathrm{cut}}$  & The typical mass scale to host a central  & [12.5, 13.7]\\
  &  $\log_{10}{M_{1}}$  & The typical mass scale for halos to host one satellite  & [13.6, 15.1] \\
   &  $\log_{10}\sigma$  & The turn on slope for central occupation  & [-2.99, 0.96] \\
  &  $\alpha$  & The power-law index for the mass dependence of the number of satellites & [0.30, 1.48]\\
   &  $\kappa$  & Parameter that defines the minimum mass to host a satellite & [0.00, 0.99]\\
   & $\alpha_\mathrm{vel, c}$ & Central velocity bias                         & [0.00, 0.61] \\
   & $\alpha_\mathrm{vel, s}$ & Satellite velocity bias                         & [0.58, 1.49] \\
   & $B_\mathrm{cent}$ & Central environment-based secondary bias                        & [-0.67, 0.20] \\
   & $B_\mathrm{sat}$ & Satellite environment-based secondary bias                       & [-0.97, 0.99] \\

\hline
Optional parameters   & $A_\mathrm{cent}$ & Central concentration-based secondary bias                        & [-0.99, 0.93] \\
   & $A_\mathrm{sat}$ & Satellite concentration-based secondary bias                       & [-1.00, 1.00] \\
   & $s$ & satellite profile bias parameter                                                   & [-0.98, 1.00] \\

\hline
\end{tabular}

\caption{The parameters used in our emulators, their physical interpretation and their range in the our training set. The optional parameters are not included in our fiducial emulator, but are explored in section~\ref{subsec:additional_extensions}. The ranges represent the bounds on the emulator training set. The cosmology bounds come from the minimum bounding box around the \textsc{AbacusSummit} cosmology grid, which is in turn sampled from 5 to 8 standard deviations beyond current constraints from the combination of CMB and large-scale structure data. The HOD bounds come from the the minimum and maximum in each parameter within the 2550 HOD training set. }
\label{tab:parameters}
\end{table*}

\subsection{Emulator leave-one-out test}
\label{subsec:emu_test}

To assess the accuracy of our emulator, we conduct leave-one-out tests at 5 test cosmologies (c000-c004, \red{shown in orange in Figure~\ref{fig:cosmo_grid}}). \red{We only test at 5 out of the 85 cosmologies due to the high computational cost of training GP models (each leave-one-out test requires re-training the emulators). We select the 5 cosmologies to span a large but plausible range in the main parameter space of interest ($\omega_\mathrm{cdm}$ vs $\sigma_8$), with the caveat that these 5 cosmologies do not span some other parameter directions, most notably $\omega_b$ and $\alpha_s$. However, we justify this choice by noting that the  small-scale RSD measurement will be most sensitive to $\omega_\mathrm{cdm}$ and $\sigma_8$, and we do not expect strong constraining power along other cosmology directions.} 

\begin{figure}
    \centering
    \hspace*{-0.3cm}
    \includegraphics[width = 3.6in]{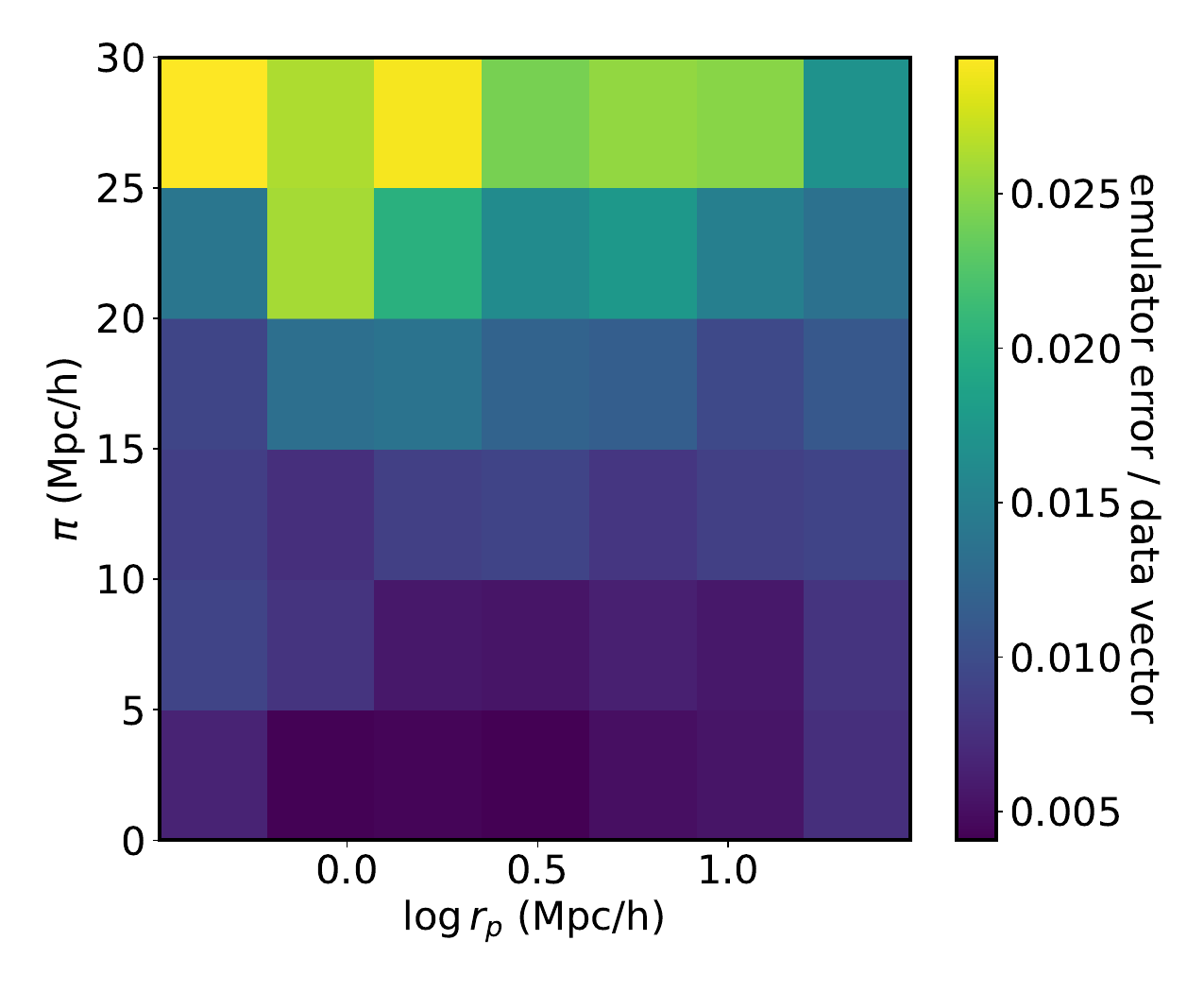}
    \vspace{-0.3cm}
    \caption{The fractional emulator error as a function of separation bins in our leave-one-out tests. We achieve sub $1\%$ error in the low $\pi$ bins, out to $r_p = 30h^{-1}$Mpc. The error increase to $<3\%$ at large $\pi$, where the data signal-to-noise is rather poor.}
    \label{fig:test_emulator_frac}
\end{figure}

\begin{figure}
    \centering
    \hspace*{-0.3cm}
    \includegraphics[width = 3.6in]{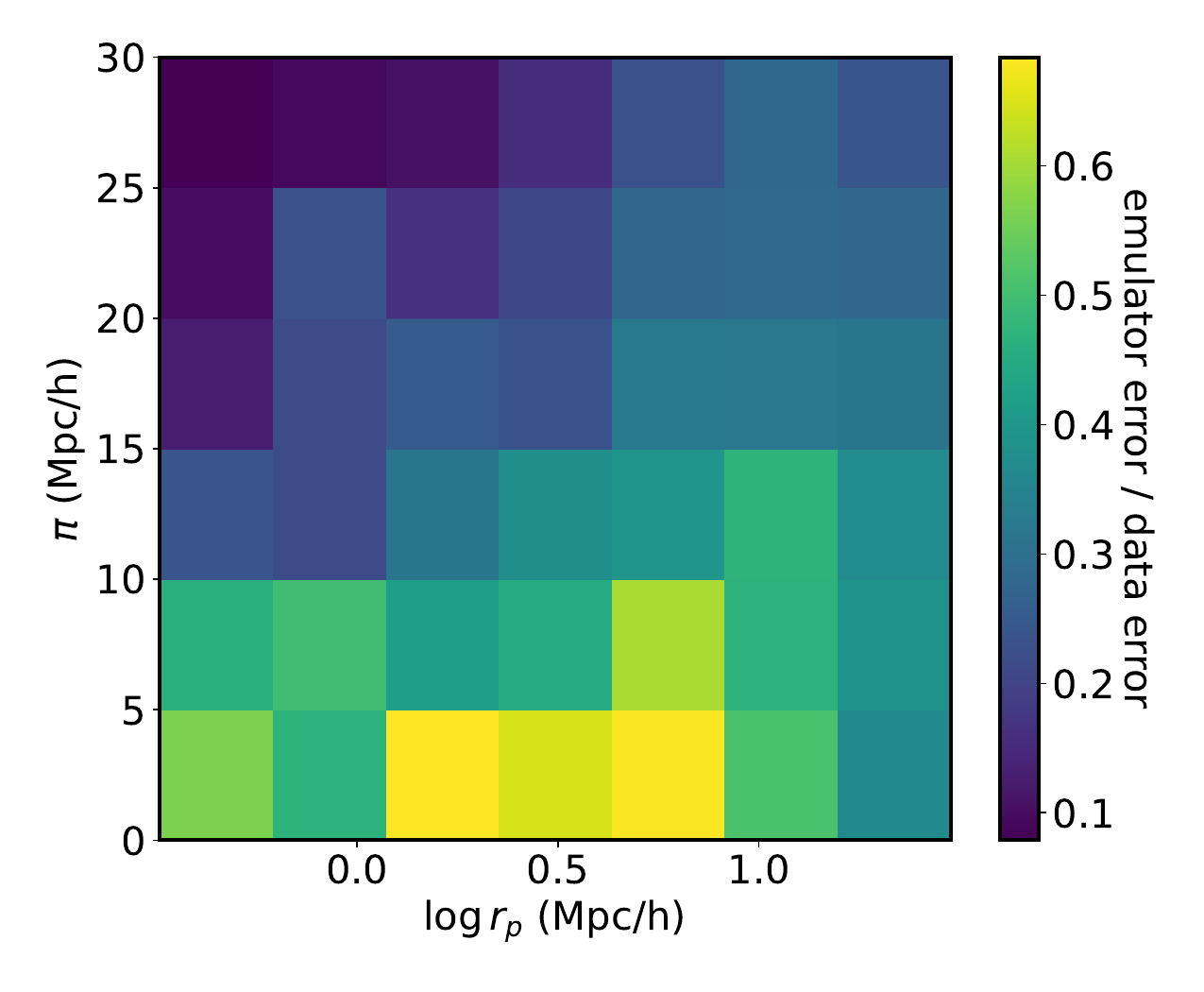}
    \vspace{-0.3cm}
    \caption{The emulator error relative to the CMASS jackknife error. We find the emulator error to be sub-dominant compared to the data uncertainty in all bins. The largest emulator error contribution comes in a few bins at low $\pi$ and $r_p$ of a few megaparsecs, where the emulator error reaches $\sim 70\%$ of the data error. However, the typical emulator error across most separation bins is around $30-40\%$ of the data error.}
    \label{fig:test_emulator_rel}
\end{figure}

For each one of the five tests, we leave out 1 cosmology and its associated HOD samples out of the training set. Once we have trained the emulator on the other 84 cosmologies, we randomly select 100 HODs from the HOD MC chain run at the left-out cosmology as the test set. We also re-run each of the 100 test HODs with 16 different random seeds and then take the average clustering prediction to suppress the shot noise. Finally, we use the emulator to make clustering predictions at the test cosmology and the 100 test HODs. We report the mean absolute fractional errors averaged across the five test cosmologies in Figure~\ref{fig:test_emulator_frac}. In Figure~\ref{fig:test_emulator_rel} we show the same errors, but normalized by CMASS data error, which is calculated as the diagonal square root of the jackknife covariance matrix.  
Figure~\ref{fig:test_emulator_frac} shows that we achieve sub-percent level accuracy for bins at low $\pi$, which are the bins with the highest measurement signal-to-noise. At larger $\pi$, the fractional error increases to $<3\%$. However, if we compare these errors to data jackknife errors, as shown in Figure~\ref{fig:test_emulator_rel}, the picture is completely different. The data error is 10 up to times larger than emulator errors at large $\pi$ bins. At low $\pi$ bins, the emulator error remains sub-dominant compared to the data error, but increases to up to $<70\%$ of the data error. The mean and median error across all bins is approximately 30--50$\%$ of the data error.

\red{
For the rest of this paper, we quote the typical error of our emulator to be approximately 30--50$\%$ of the data error. This is a zeroth-order approximation of the true emulator error, which varies both as a function of cosmology and HOD. In our tests, the typical error of the emulator as a fraction of data error for each of the 5 test cosmologies (c000-c004) are 0.09, 0.45, 0.29, 0.44, 0.15, respectively. As expected, our emulator performs best at Planck cosmology (c000) at the center of the cosmology grid. The emulator does fairly well at c004, which has a lower than Planck $\sigma_8$. The errors are the largest at c001 and c003, which have both a deviant $\sigma_8$ and $\omega_\mathrm{cdm}$. Nevertheless, the typical errors across all 5 test cosmologies are less than 50$\%$ of the data error. Thus, within the most likely region of cosmology parameter space, our subsequent analysis should be limited by data error, not emulator error. However, the number of tests we conduct is limited by computation time, and we caution that our emulators could report larger errors in untested regions of cosmology space. Finally, We note for future studies that the more precise thing to do would be do characterize the emulator error as a function of cosmology and HOD and use the estimated emulator error at the best-fit cosmology and HOD to compute the final covariance matrix.}


\subsection{Covariance matrix}
\label{subsec:cov_bias}
We define our log-likelihood function as 
\begin{equation}
       \chi^2_{\xi}  = (\xi_{\mathrm{emulator}} - \xi_{\mathrm{data}})^T \boldsymbol{C}^{-1}(\xi_{\mathrm{emulator}} - \xi_{\mathrm{data}}),
       \label{equ:chi2}
\end{equation}
where the $\xi_{\mathrm{emulator}}$ is the emulator predicted $\xi(r_p, \pi)$ and $\xi_\mathrm{data}$ is the CMASS measurement. $\boldsymbol{C}$ is the covariance matrix. 

Properly accounting for covariance contributions due to data and modeling systematics is essential in obtaining robust constraints on cosmology and HOD. For this analysis, we decompose the covariance matrix into three individual terms
\begin{equation}
    \boldsymbol{C} = \boldsymbol{C}_\mathrm{data} + \boldsymbol{C}_\mathrm{emulator} + \boldsymbol{C}_\mathrm{mock,sample}.
    \label{equ:cov}
\end{equation}
The first term $\boldsymbol{C}_\mathrm{data}$ stands for the data covariance contribution, coming mostly from the data sample variance due to limited sample size. We briefly discussed the data covariance matrix in section~\ref{sec:data}, and visualized the corresponding correlation matrix in the right panel of Figure~\ref{fig:xi_boss}. Again, we compute $\boldsymbol{C}_\mathrm{data}$ through jackknife resampling over 400 sub-regions in the CMASS sample. We choose the number of jackknife regions to be large enough to yield a well-conditioned covariance matrix, and small enough to give sufficient volume to each jackknife region. We validate our jackknife covariance matrix against a mock-based covariance matrix in Appendix~\ref{sec:cov_comapre}, leveraging the 25 random phase simulation boxes. 

The $\boldsymbol{C}_\mathrm{emulator}$ term corresponds to the uncertainty due to emulator error, which we have described in Section~\ref{subsec:emu_test}. We compute this term using the 100 test HODs at 5 test cosmologies generated in the leave-one-out tests we described in the previous section. Figure~\ref{fig:test_emulator_rel} shows the relative squareroot magnitude of $\boldsymbol{C}_\mathrm{emulator}$ compared to $\boldsymbol{C}_\mathrm{data}$, suggesting that in general, the covariance contribution due to emulator errors is roughly 10--30$\%$ that of the CMASS data covariance matrix. \red{While, we have only estimated the emulator error using a relatively small number of test samples, this nevertheless indicates that the posterior uncertainty of this analysis will be limited by data, not the modeling uncertainties. In the case of DESI, the data sample variance will be significantly smaller than CMASS. Thus, it becomes critical to further improve the emulator accuracy and conduct a more careful characterization of the emulator error. We defer a more detailed discussion of DESI prospects to Section~\ref{subsec:desi}.}


The $\boldsymbol{C}_\mathrm{mock,sample}$ term corresponds to the sample variance contribution due to finite simulation volume. Given that a single \textsc{AbacusSummit} base volume is approximately 6 times that of the CMASS sample, the mock sample variance should constitute at most a $\sim 16\%$ contribution compared to the data covariance, making it a minor component of the total covariance. \red{However, looking forward to DESI, we expect the data sample variance to be a few times smaller than that of CMASS. In Appendix~\ref{sec:phase}, we demonstrate a phase correction routine that would further reduce the sample variance in the model. Such phase comparison is not important for the accuracy of the CMASS dataset.}

Finally, we note the omission of the shot noise contribution to the overall covariance matrix associated with the random nature of the HOD realization. We find that the large simulation volume combined with averaging over 16 HOD realizations for each training and test sample reduced the shot noise contribution to the overall covariance matrix down to approximately $1\%$, insignificant given the data uncertainties in this specific analysis. 

\subsection{Nested sampling and prior setup}
\label{subsec:sampling}
So far, we have described our emulator model and its training, the likelihood model and its associated covariances and biases. In this section, we describe the sampler we employ to sample the likelihood surface and derive posterior constraints. 

Specifically, we use the \textsc{dynesty} nested sampler \citep{2018Speagle, 2019Speagle}. While being able to sample the posterior space more efficiently than an Markov Chain Monte Carlo sampler, nested sampling codes such as \textsc{dynesty} also compute the Bayesian evidence, 
\begin{equation}
    \mathcal{Z} = P(D|M) = \int_{\Omega_{\Theta}} P(D|\Theta, M)P(\Theta|M)d\Theta.
    \label{equ:evidence}
\end{equation}
where $M$ represents the predictive model, $D$ represents the observed $\xi$, and $\Theta$ represents the cosmology and HOD model parameters. $P(D|\Theta, M)$ is the likelihood function, and $P(\Theta|M)$ is the Bayesian prior. The evidence can simply be interpreted as the marginal likelihood of the data given the model, and serves as an important metric in Bayesian model comparisons. In our \textsc{dynesty} runs, we use 500 live points and a random walker sampler. The stopping criterion is set to $d\log\mathcal{Z} > 0.01$. 

\red{
For our fiducial analysis, we use broad multi-variate Gaussian priors with bounds as our prior. We construct the cosmology and HOD priors separately. 
For the cosmology parameters, we choose a weakly informative Gaussian prior to incorporate some existing constraints on parameters that our small-scale RSD signal is insensitive to, at least at current level of signal-to-noise in the data. These parameters include, for example $\omega_b, \alpha_s, w_0, w_a$. Thus, for the cosmology paramters, we first define a Gaussian prior with mean at fiducial Planck cosmology and covariance calculated from the cosmology grid, which was set up around Planck + LSS 3-8$\sigma$ constraints.
Then for the cosmology parameters of interest, such as $\omega_\mathrm{cdm}$ and $\sigma_8$, we inflate their prior width in order to ensure their posteriors are not prior-dominated. Specifically, we do a grid search where we progressively inflate the Gaussian prior width by an arbitrary amount, until the posterior constraints fully converge. We find that an inflation of $\times 1.5$ in the prior width ensures such convergence.
In addition to the Gaussian priors, we also impose a generous bounding box to safeguard against unphysical parameter values, with edges that are wider than those listed in Table~\ref{tab:parameters}.
In Appendix~\ref{subsec:flat_prior}, we also repeat our fiducial analysis with a flat prior and demonstrate that our results are robust against prior choices.}

\red{
For the HOD prior, we define a separate multivariate Gaussian prior with mean and variance computed from the 2550 HOD training set. We also inflate the HOD prior width by $50\%$ to make the prior less informative. Again, we apply an additional bounding box that is exterior to the prior ranges quoted in Table~\ref{tab:parameters} to safeguard the sampler from reaching implausible parameter values or cross limits imposed by HOD parameter definitions, e.g. $\kappa > 0$ and $\alpha_\mathrm{vel, c} > 0$. }

\subsection{Recovery test}
Before we apply our inference pipeline to data, we first conduct a recovery test where we generate a mock data vector at known cosmology and HOD and use our inference pipeline to attempt to recover the correct underlying cosmology and HOD from the mock data vector.

\begin{figure*}
    \centering
    \hspace*{-0.3cm}
    \includegraphics[width = 7.1in]{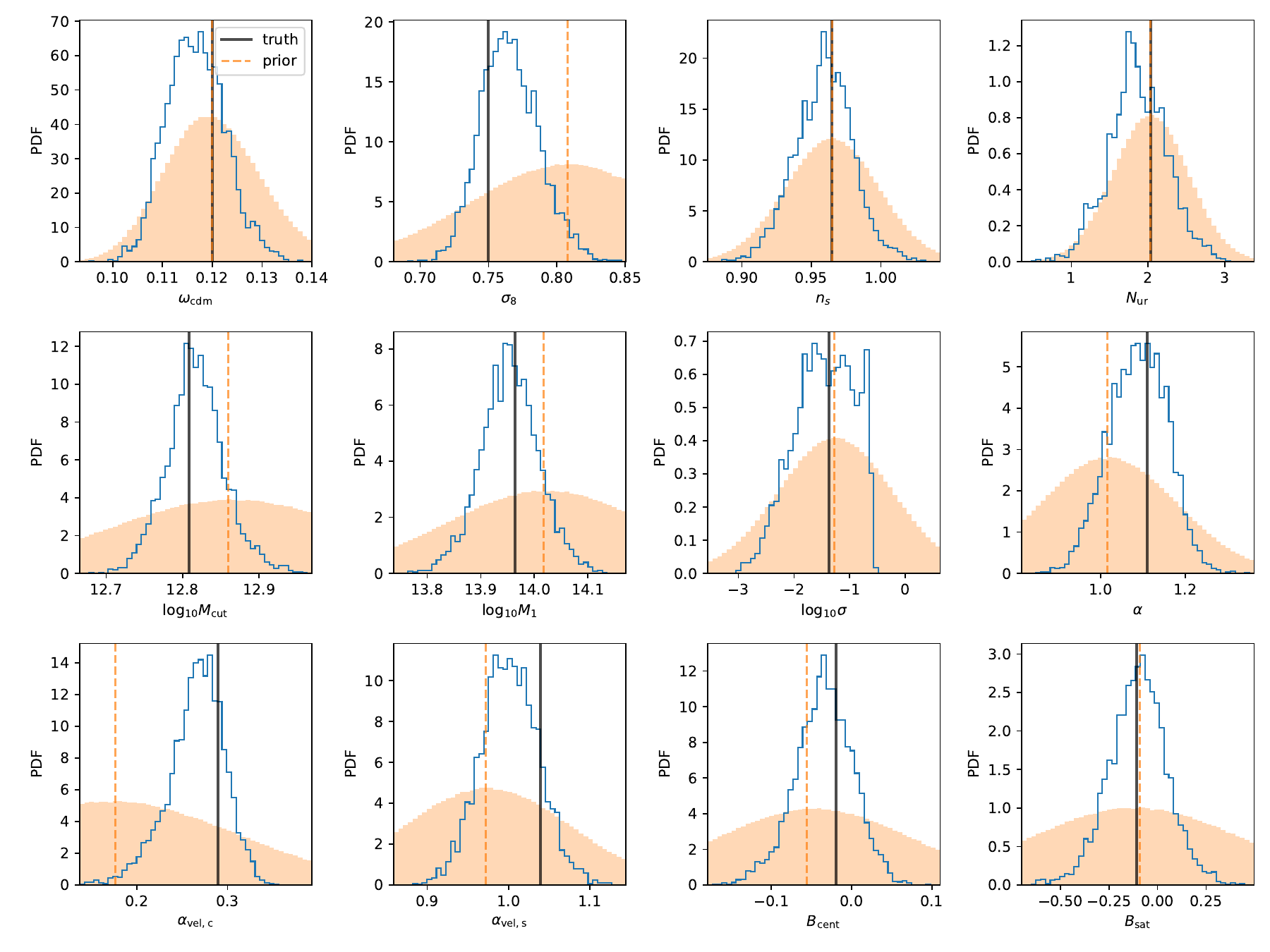}
    \vspace{-0.3cm}
    \caption{The mock recovery test conducted on a mock RSD data vector generated at c004 (lower $\sigma_8$) and a fiducial HOD. The true underlying cosmology and HOD parameters of the mock data vector are marked by the vertical black lines. The orange histograms and lines denote the prior distribution and prior mean. The blue histogram showcase the posterior constraints. We obtain good recovery ($<1\sigma$) on all parameters shown. We do not show $\omega_b, \alpha_s, n_0, n_a$ as our small-scale RSD data vector has little constraining on these parameters. We have also omitted HOD parameter $\kappa$, which controls the cutoff mass for satellite galaxy occupation, as it is not interesting and does not impact small-scale RSD significantly for a high-mass sample like CMASS.}
    \label{fig:recovery}
\end{figure*}

For our mock data vector, we pick cosmology c004, which has a lower $\sigma_8  = 0.75$, and a test HOD as shown by the black vertical lines in Figure~\ref{fig:recovery}. The HOD is picked from the MC HOD chain at cosmology c004, so we know that the resulting mock data vector resembles the real data vector. We generate the mock data vector $\xi(r_p, \pi)$ using the same separation binning as the CMASS data vector (Figure~\ref{fig:xi_boss}). We average over 16 HOD realizations to reduce shot noise. Then we train our emulator using the other 84 cosmologies (excluding c004) and 2550 HODs. Finally, we run the nested sampling inference pipeline with the combined covariance matrix and the inflated ($\times 1.5$) multivariate Gaussian prior. We do not invoke the phase correction as the mock data vector is generated at fixed phase ph000. The resulting marginalized 1D constraints are summarized in Figure~\ref{fig:recovery}, where the black lines denote the true parameter values of the mock data vector, and the orange histograms and lines denote the prior PDF and prior mean. The blue histograms show the marginalized 1D posteriors. We have omitted four cosmological parameters $\omega_b, \alpha_s, n_0, n_a$ as our small-scale RSD data vector has little constraining power on these parameters. The recovery on these parameters are entirely prior dominated. We have also omitted HOD parameter $\kappa$, which controls the cutoff mass for satellite galaxy occupation, as it is not interesting and does not impact small-scale RSD significantly for a high mass sample like CMASS. 

Figure~\ref{fig:recovery} shows that we achieve good recovery on all parameters, within $1\sigma$ uncertainty. It is particularly important that we successfully recover a low $\sigma_8$, as that is of top scientific interest in this analysis. For the other cosmology parameters shown, we obtain a modest amount of constraining power on two other structural growth parameters $\omega_\mathrm{cdm}$ and $n_s$, but little constraining power on $N_\mathrm{ur}$. For the HOD parameters, we find excellent recovery and strong constraining power on all parameters. The only parameter that shows slightly worse recovery is satellite velocity bias $\alpha_\mathrm{vel, s}$, which the posterior mean deviates from the truth by about $1\sigma$. This discrepancy is not reproduced when using other cosmology or HODs as true values. 

\begin{figure}
    \centering

    \begin{subfigure}[t]{0.5\textwidth}
    \includegraphics[width=3in]{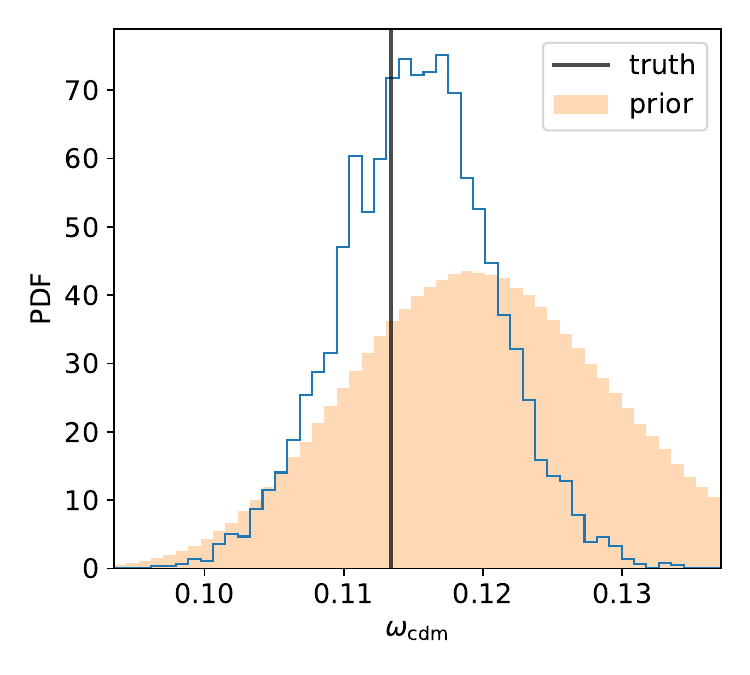} 
    \vspace{-0.3cm}
    \caption{Recovery c001}
    \end{subfigure}
    
    \begin{subfigure}[t]{0.5\textwidth}
    \includegraphics[width=3in]{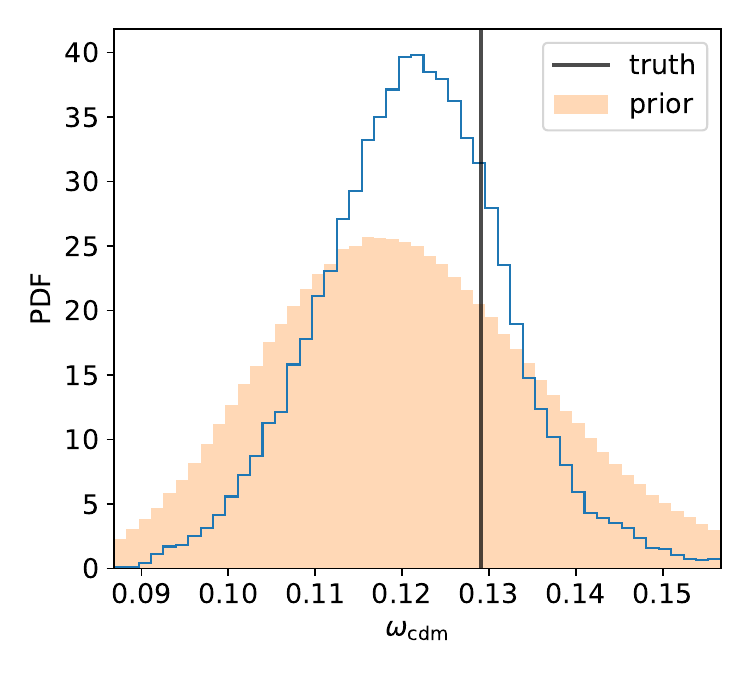} 
    \vspace{-0.3cm}\caption{Recovery c003}
    \end{subfigure}
    
    \caption{The mock recovery test conducted on c001 (top) and c003 (bottom) cosmologies to test the recovery of $\omega_\mathrm{cdm}$. Both tests find posteriors statistically consistent with the truth values.}
    \label{fig:rec_omegam}
\end{figure}

\mage{Additionally, we repeat the same test for cosmologies c001 and c003 to test the recovery of the mass density parameter $\omega_\mathrm{cdm}$. These two cosmologies share the same fiducial values for all cosmological parameters except for $\omega_\mathrm{cdm}$. For brevity, we only show the $\omega_\mathrm{cdm}$ panel of these two tests in Figure~\ref{fig:rec_omegam}, with the first and second panel showing the recovery in c001 and c003, respectively. Both tests find posteriors that are statistically consistent with the truth. }

\section{Results}
\label{sec:results}
In this section, we describe the cosmology and HOD constraints in our fiducial analysis of the CMASS small-scale RSD measurement and their broader implications. 

\subsection{Fiducial analysis setup}
While the emulator was trained over the 8-parameter $w_0w_a$CDM+$N_\mathrm{eff}$+$\alpha_s$ parameter space, for the fiducial analysis, we fix the dark energy equation-of-state parameters $w_0 = -1$ and $w_a = 0$ and limit the cosmology sampling to $\Lambda$CDM+$N_\mathrm{eff}$+$\alpha_s$ parameter space. We fix $w_0$ and $w_a$ because the small-scale RSD is not sensitive to these two parameters and the fact that our analysis is not yet set up to account for modulations in radial distances as a function of these two parameters. 

To summarize the inference pipeline as described in the last section, we first run MC chains in HOD parameter space at each of the 85 cosmologies, from which we select a 85 cosmology and 2550 HOD training set (section~\ref{subsec:method_emulator}). We train the GP-based emulator model on said training set and then fit the emulator model to the CMASS RSD measurement using a nested sampler. For the likelihood calculation, we adopt broad Gaussian priors (section~\ref{subsec:sampling}) and a covariance matrix combining contributions from the data, the emulator errors, and mock sample variance (section~\ref{subsec:cov_bias}). We additionally apply a phase correction term to convert our model prediction from fixed-phase to mean-phase (Equation~\ref{equ:phase_correct}), while also reducing the sample variance contribution. 

For our fiducial analysis, we train our emulator using a standard 5-parameter HOD extended with velocity bias and environment-based secondary bias, for a total of 9 HOD parameters. The HOD parameters are summarized in the HOD section of Table~\ref{tab:parameters}. However, for the purpose of comparison, we run two sets of inferences with two different HOD models, one with environment-based bias and one without. This is to test our cosmology inferences' robustness to HOD modeling, but also more specifically to address the interesting question of whether galaxy environment-based bias, or ``galaxy assembly bias'', affects cosmological inference. As described previously in Section~\ref{subsubsec:hod_Ext}, the choice of velocity bias and environment-based secondary bias as the HOD extensions for the fiducial analysis is motivated by HOD analyses of LRG mocks selected from hydrodynamical simulations \citep[]{2022Yuan, 2020Hadzhiyska, 2020Xu, 2021Xu, 2021Delgado}.

Throughout the figures presented in this section, we use blue contours to indicate fits without environment-based bias and red contours to indicate fits with environment-based contours. Table~\ref{tab:fiducial_fits} summarizes the two fiducial fits, displaying goodness-of-fits, Bayesian evidence, and the constraints on cosmology and HOD parameters. Later on in section~\ref{subsec:additional_extensions}, we consider the effects of additional HOD parameters, including one-halo term modulations, and concentration-based secondary bias. 

In terms of goodness-of-fit, both fits achieve reasonably good fits on data, but the inclusion of the two environment-based bias parameters does improve both the $\chi^2$/d.o.f. and the log evidence, as shown in Table~\ref{tab:fiducial_fits}. The d.o.f. presented are the effective degrees-of-freedom, accounting for parameter degeneracies and unconstrained parameters such as $w_0$ and $w_a$. Specifically, we follow the derivations of \citet{2019Raveri}. We have also applied corrections to the presented $\chi^2$ and $\log\mathcal{L}$ values following \citet{2007Hartlap}. 

\begin{table*}
\centering
\begin{tabular}{llccc}
\hline
& Parameters & without environment-based bias    & with environment-based bias & Prior \\
\hline
Goodness-of-fit   & $\log \mathcal{L}$              & -18.9   & -16.4  & -\\
   & d.o.f.                                         & 32.8 & 31.0 & -\\
   & $\chi^2$/d.o.f.                                & 1.15 & 1.06 & -\\
   & $\log \mathcal{Z}$                             & -40.9 & -38.8 & -\\
\hline

Cosmology constraints ($68\%$ C.L.)& $\omega_{b}$   & $0.0221\pm 0.0008$  & $0.0226\pm 0.0008$  & $0.0224\pm 0.0009$\\
  & $\omega_\mathrm{cdm}$                           & $0.115\pm 0.004$ &  $0.115\pm 0.004$ & $0.12\pm 0.01$\\
  &  $\sigma_8$                                     & $0.756\pm 0.016$ &  $0.762\pm 0.024$  & $0.81\pm 0.07$\\
   &  $n_{s}$                                       & $0.94\pm 0.02$ &  $0.95\pm 0.02$ & $0.96\pm 0.03$\\
   &  $\alpha_{s}$                                  & $-0.007\pm 0.015$ &  $-0.004\pm 0.015$  & $0.00\pm 0.02$\\
  &  $N_{\text{ur}}$                               & $1.7\pm 0.3$ &  $1.8\pm 0.3$ & $2.0\pm 0.5$\\
  
\hline 
  HOD constraints ($68\%$ C.L.)&  $\log_{10}{M_\mathrm{cut}}$   & $12.79\pm 0.03$  & $12.77\pm 0.03$ & $12.86\pm 0.17$\\
  &  $\log_{10}{M_{1}}$                             & $13.94\pm 0.05$ &  $13.94\pm 0.06$ & $14.02\pm 0.19$\\
   &  $\log_{10}\sigma$                             & $-1.7\pm 0.6$  &  $-1.6\pm 0.5$ & $-1.3\pm 1.0$\\
  &  $\alpha$                                       & $1.10\pm 0.06$ & $1.05\pm 0.06$ & $1.01\pm 0.17$\\
   &  $\kappa$                                      & $0.4\pm 0.2$ & $0.5\pm 0.3$ & $0.5\pm 0.4$\\
   & $\alpha_\mathrm{vel, c}$                       & $0.29\pm 0.03$ &  $0.23\pm 0.04$ & $0.18\pm 0.14$\\
   & $\alpha_\mathrm{vel, s}$                       & $1.00\pm 0.04$    &  $0.99\pm 0.04$ & $0.97\pm 0.10$\\
   & $B_\mathrm{cent}$                              &  -   &  $0.02\pm 0.04$ & $-0.05\pm -0.13$\\
   & $B_\mathrm{sat}$                               &  -  &  $-0.36\pm 0.19$ & $-0.1\pm 0.6$\\
\hline
\end{tabular}

\caption{The parameter constraints from our fiducial fits on CMASS small-scale RSD and the associated priors. The first two data columns summarize the two fiducial fits, with and without environment-based bias. The last column lists the prior spread. The first horizontal block lists the goodness-of-fit parameters. $\mathcal{L}$ denotes the maximum likelihood. d.o.f. denotes the effective degrees-of-freedom, calculated following \citet{2019Raveri}. $\mathcal{Z}$ denotes the Bayesian evidence. All error bars represent $68\%$ C.L.}
\label{tab:fiducial_fits}
\end{table*}

\begin{figure*}
    \centering
    \hspace*{-0.3cm}
    \includegraphics[width = 7.5in]{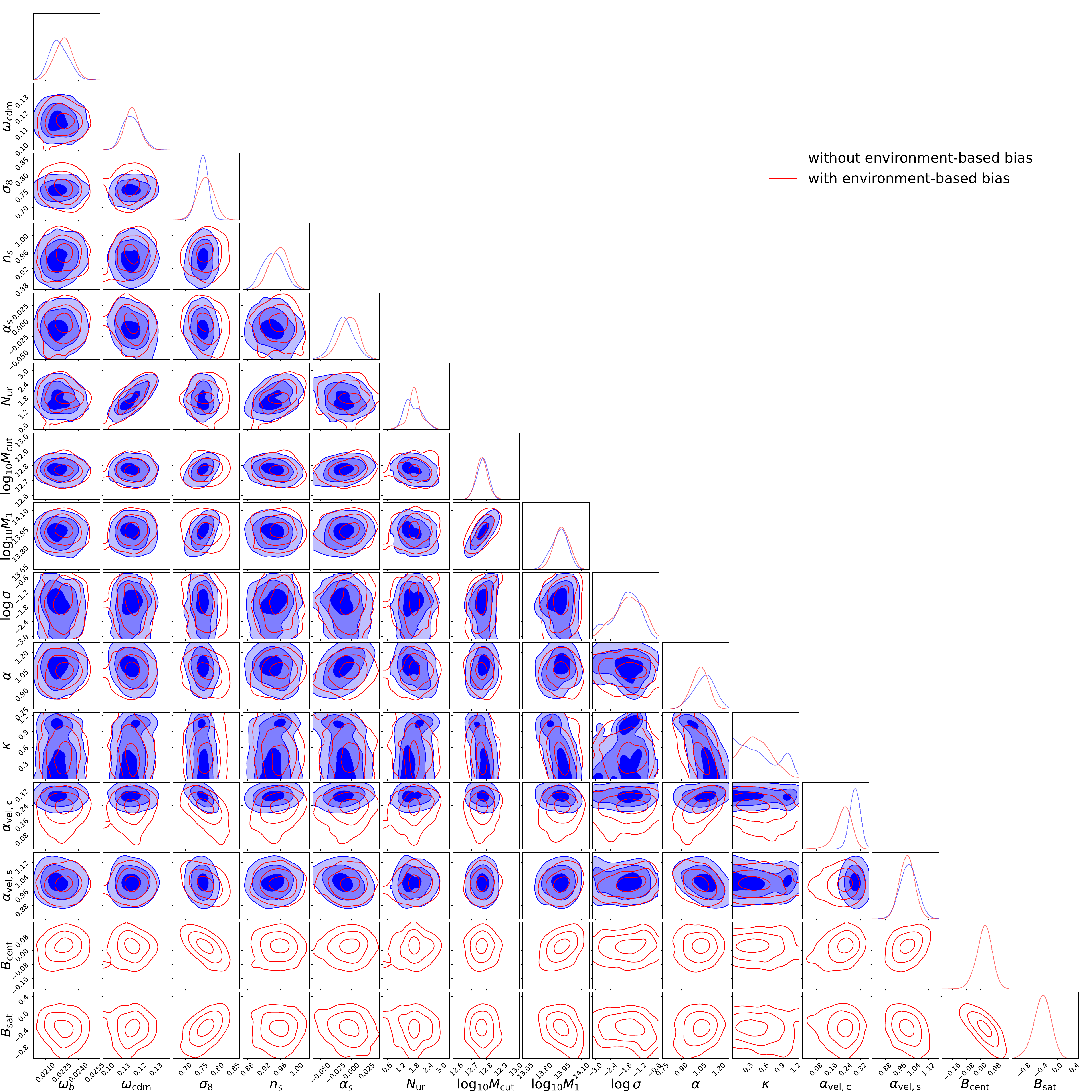}
    \vspace{-0.3cm}
    \caption{The 2D marginalized constraints on cosmology and HOD parameters from the CMASS small-scale RSD measurement. The red and blue contours showcase the constraints with and without enabling environment-based secondary biases in the HOD, respectively. The contours represent the $1,2,3\sigma$ constraints. This plot most-importantly shows that while the inclusion of environment-based bias somewhat loosens the constraining power on certain cosmological parameters, it does not bias those constraints. For readability, we show marginalized constraints on subsets of parameters in following figures. }
    \label{fig:corner_full}
\end{figure*}

Figure~\ref{fig:corner_full} shows the full 2D marginalized cosmology and HOD constraints. The red and blue lines showcase the constraints with and without enabling environment-based secondary biases in the HOD, respectively. The contours represent the $1,2,3\sigma$ constraints. 
Because of the high dimensionality, this figure is not particularly helpful in visualizing the exact constraints on each parameter. For that reason, we zoom in on marginalized constraints for subsets of parameters in subsequent figures. However, Figure~\ref{fig:corner_full} serves two important purposes. First of all, it shows that the inclusion of the environment-based secondary biases does not significantly bias the cosmological inference, but it does seem to loosen the constraints somewhat. Second, it showcases the degeneracies between all parameter combinations. The cross-correlations between HOD extensions and cosmology parameters are particularly interesting. 

Specifically, the clustering amplitude $\sigma_8$ appears positively correlated with the HOD mass parameters $\log M_\mathrm{cut}$ and $\log M_1$ but negatively correlated with the satellite power law index $\alpha$. This makes sense as a trade-off between matter clustering and galaxy bias. More interesting, $\sigma_8$ correlates with both HOD extensions. In terms of velocity bias, $\sigma_8$ negatively correlates with central velocity bias $\alpha_\mathrm{vel, c}$. This makes sense as a higher $\sigma_8$ yields deeper gravitational potential wells, thus requiring less galaxy velocity bias. In terms of secondary biases, $\sigma_8$ is negatively correlated with central environment-based bias $B_\mathrm{cent}$ but positively correlated with satellite environment-based bias $B_\mathrm{sat}$. The negative correlation between $\sigma_8$ and $B_\mathrm{cent}$ indicates that at smaller matter clustering amplitude, centrals preferentially occupy halos in less dense environments (Positive $B$ indicates preference for less dense environments). We do not have a clean explanation for this effect, but we speculate it is likely due to correlations with other HOD parameters as well as cosmology. It is worth noting that $B_\mathrm{cent}$ is relatively well-constrained at 0, and the negative correlation is small in amplitude. The positive correlation between $\sigma_8$ and $B_\mathrm{sat}$ makes sense as it indicates satellites preferentially occupy halos in denser environments when $\sigma_8$ is low, consistent with the idea of a trade-off between galaxy bias and underlying matter clustering. 

\subsection{Cosmology constraints}
\begin{figure*}
    \centering
    \hspace*{-0.3cm}
    \includegraphics[width = 7.2in]{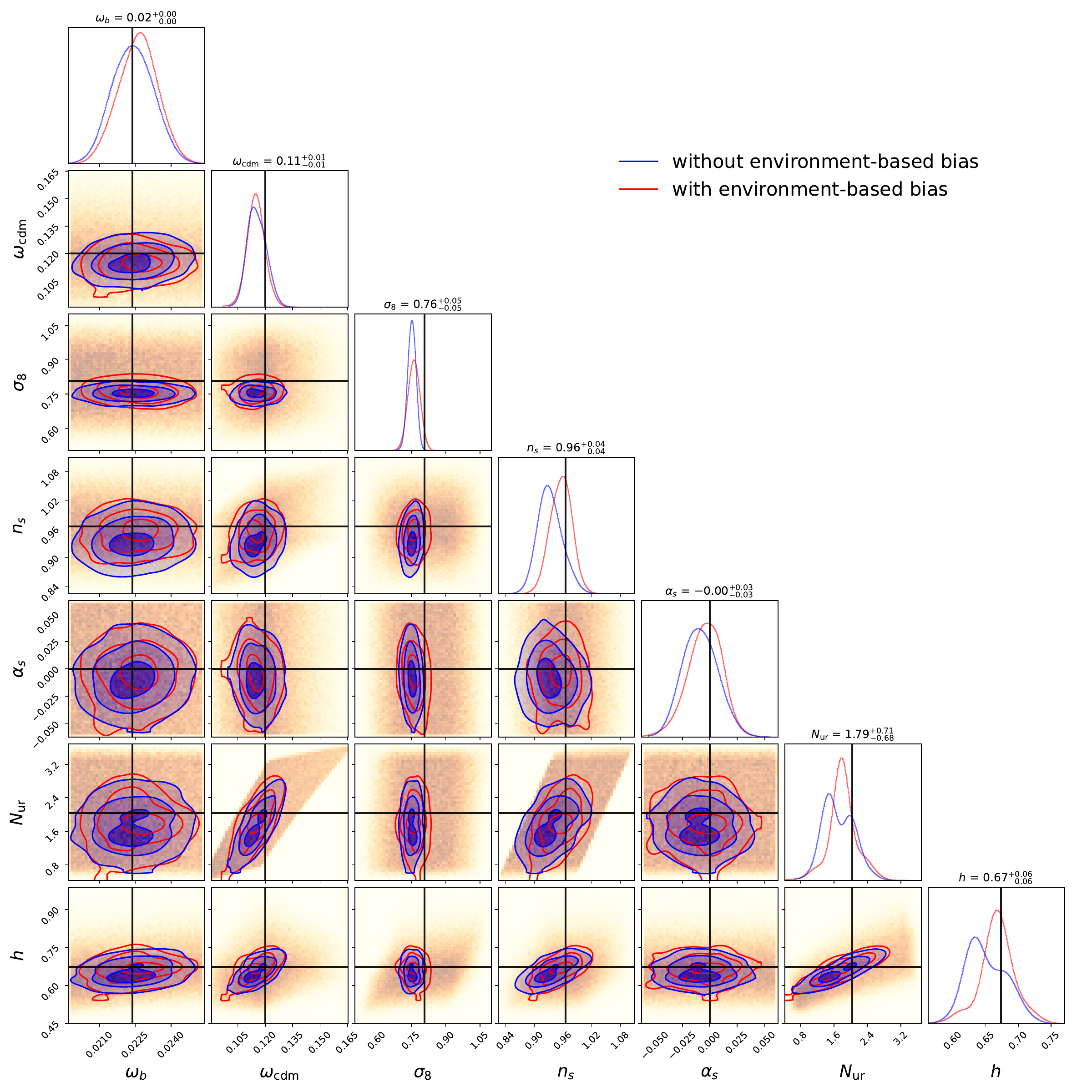}
    \vspace{-0.3cm}
    \caption{The 2D marginalized constraints on cosmology parameters from the CMASS small-scale RSD measurement. The red and blue contours showcase the 1-3 $\sigma$ constraints with and without enabling environment-based secondary biases in the HOD, respectively. The underlying heatmaps show the prior volume. The edges of the panels showcase the bounding box applied to the priors. The titles on the 1D constraints show the posterior mean and $2\sigma$ constraints in the with environment-based bias case, ($95\%$ C.L.). The black lines denote fiducial Planck 2018 values. }
    \label{fig:corner_cosmo}
\end{figure*}

In this and the following subsections, we examine the marginalized cosmology and HOD parameter constraints. Figure~\ref{fig:corner_cosmo} showcases the marginalized cosmology constraints within a $\Lambda$CDM+$N_\mathrm{eff}$+$\alpha_s$ model plus the inferred Hubble parameter $h$ (fixed acoustic scale). The $2\sigma$ ($95\%$ C.L.) marginalized constraints are displayed in the title and collected in Table~\ref{tab:fiducial_fits}. The red and blue contours correspond to the 1--3$\sigma$ posterior constraints with and without environment-based secondary biases, respectively. The underlying orange heatmap shows the prior volume. The edges of the panels showcase the bounding box applied to the priors. The black lines correspond to the fiducial Planck values.

\red{Figure~\ref{fig:corner_cosmo} and Table~\ref{tab:fiducial_fits} reveal which cosmology parameters the CMASS small-scale RSD measurements can actually inform. We discuss each parameter in order of increasing constraining power. First, we see that the small-scale RSD does not inform the baryon density $\omega_b$, which we expected. 
Next, the data appear to be weakly constraining of three parameters: $n_s, \alpha_s$, and $N_\mathrm{ur}$. The posterior constraints for these three parameters are 25-40$\%$ tighter than their respective priors.
As for the spectral index $n_s$ and its running $\alpha_s$, the exclusion of environment-based bias shifts the spectral index $n_s$ constraints lower by approximately $1\sigma$, suggestive of a degeneracy between spectral shape and environment-based bias. This makes sense as both the spectral shape and environment-based bias module the small-scale clustering. 
The data also weakly constrain the effective number of relativistic species $N_\mathrm{ur}$. This makes sense as relativistic neutrinos smooth out the local density peaks, thus affecting the small-scale RSD. 
 The strongest constraint we obtain are with the dark matter density $\omega_\mathrm{cdm}$ and the power spectrum amplitude $\sigma_8$, as they most directly contributes to the observed clustering on small scales. The posterior constraints for these two parameters are more than $60\%$ tighter than their respective priors. We highlight these constraints as the main cosmology result of this analysis:}
\begin{align}
    \sigma_8 & = 0.756^{+0.017}_{-0.016} \ \ (68\%,\  \mathrm{without\ environment\text{-}based\ bias}),\nonumber \\
    \sigma_8 & = 0.762^{+0.024}_{-0.024} \ \ (68\%,\  \mathrm{with\ environment\text{-}based\ bias}),
    \label{equ:sigma_8}
\end{align}
and 
\begin{align}
    \omega_\mathrm{cdm} & = 0.115_{-0.004}^{+0.004}\ \ (68\%,\  \mathrm{without\ environment\text{-}based\ bias}),\nonumber \\
    \omega_\mathrm{cdm} & = 0.115_{-0.004}^{+0.005}\ \ (68\%,\  \mathrm{with\ environment\text{-}based\ bias}).
    \label{equ:omega_cdm}
\end{align}
Compared to the Planck best-fit values shown in black lines, our best-fit $\sigma_8$ value deviates by 3$\sigma$ without environment and $2.3\sigma$ with environment. This adds to a series of analyses of late-time clustering observables that also found $\sigma_8$ to be lower than Planck. We discuss this in greater detail in section~\ref{subsec:compare_sigma8}. 

\begin{figure}
    \centering
    \hspace*{-0.3cm}
    \includegraphics[width = 3.6in]{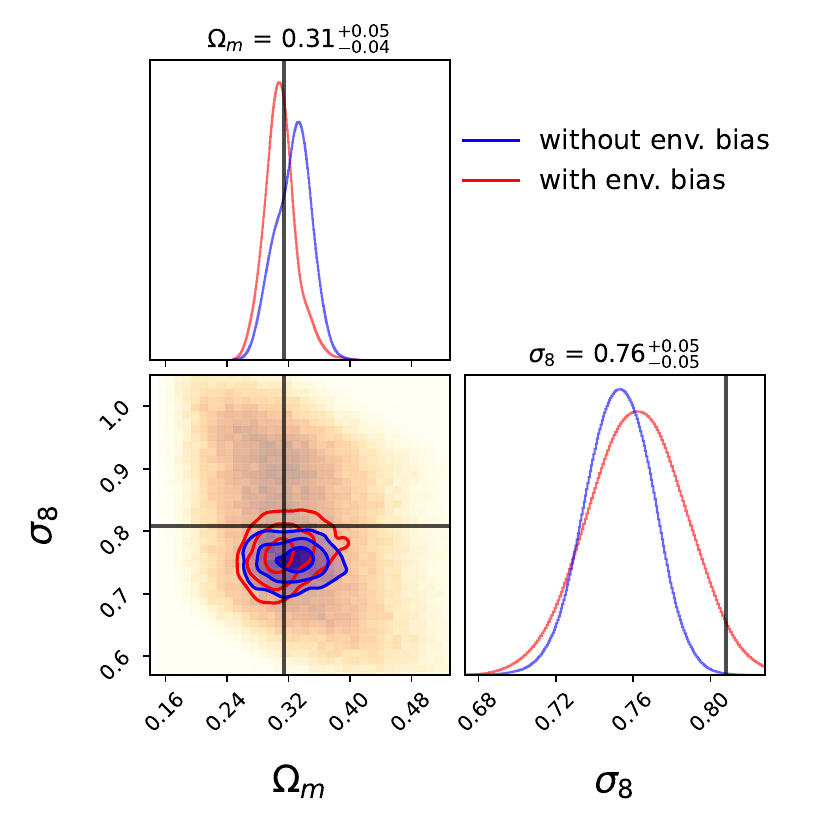}
    \vspace{-0.3cm}
    \caption{The marginalized constraints in matter density parameter $\Omega_M$ and $\sigma_8$, again showing both with and without environment-based bias. The orange heatmap shows the prior distribution, but for ease of visualization, the panel edges no longer respect the prior bounds. The title of the 1D panels correspond to the $95\%$ constraints in the with environment-based bias case. }
    \label{fig:omegam_sigma8}
\end{figure}

For completeness, we also present our constraints on a few derived parameters not listed in Table~\ref{tab:fiducial_fits}. First, we have matter density parameter $\Omega_M = (\omega_b + \omega_\mathrm{cdm})/h^2$:
\begin{align}
    \Omega_M & = 0.328^{+0.027}_{-0.020} \ \ (68\%,\  \mathrm{without\ environment\text{-}based\ bias}),\nonumber \\
    \Omega_M & = 0.309^{+0.017}_{-0.018} \ \ (68\%,\  \mathrm{with\ environment\text{-}based\ bias}).
    \label{equ:Omega_M}
\end{align}
Figure~\ref{fig:omegam_sigma8} showcases the marginalized constraints of $\Omega_M$ and $\sigma_8$, compared to the Planck values marked in black lines and the prior volume shown in the orange heatmap. The best-fit $\Omega_M$ is consistent with Planck to within $1\sigma$. 

Some RSD studies prefer to quote growth of structure constraints with the parameter combination $f\sigma_8$, where f is the growth rate of structure obtained from linear perturbation theory, $f = d \ln D/d \ln a$, where $D$ is the positive growth factor and $a$ is the scale factor. \red{To compute $f$, we first use the publicly available \textsc{colossus} code \citep[][]{2018Diemer} to analytically derive $D(a)$ for uniformly sampled cosmologies from our posterior chains. Then we compute $f\sigma_8$ constraints at $z = 0$:}
\begin{align}
    f\sigma_8(z=0) & = 0.407^{+0.018}_{-0.023} \ \ (68\%,\  \mathrm{without\ environment\text{-}based\ bias}),\nonumber \\
    f\sigma_8(z=0) & = 0.398^{+0.021}_{-0.018} \ \ (68\%,\  \mathrm{with\ environment\text{-}based\ bias}).
    \label{equ:fsigma8}
\end{align}
More commonly, the $f\sigma_8$ constraints are quoted at the effective redshift of the sample $z_\mathrm{eff}$. We derive and compare our $f\sigma_8(z_\mathrm{eff})$ constraints to Planck and other late-time clustering studies in section~\ref{subsec:compare_sigma8}.
The equivalent cosmological parameter combination to $f\sigma_8$ probed by large-scale lensing probes is $S_8 = \sqrt{(\Omega_M/0.3)}\sigma_8$. We derive marginalized $S_8$ constraints:
\begin{align}
    S_8 & = 0.786^{+0.031}_{-0.035} \ \ (68\%,\  \mathrm{without\ environment\text{-}based\ bias}),\nonumber \\
    S_8 & = 0.771^{+0.041}_{-0.037} \ \ (68\%,\  \mathrm{with\ environment\text{-}based\ bias}).
    \label{equ:S8}
\end{align}
The fiducial Planck constraints $S_8 = 0.832 \pm 0.013$ ($68\%$ TT,TE,EE+lowE+lensing) for comparison. 
Both our inferred $f\sigma_8$ and $S_8$ are mildly lower than Planck. However, because our $\Omega_M$ constraint is consistent with Planck, the discrepancy in these two parameter combinations come purely from a potential tension in $\sigma_8$. 

\red{Comparing the cosmology constraints with and without environment-based biases, we find the two sets of constraints to be broadly consistent. This suggests that the introduction of environment-based bias does not significantly bias the cosmology inference, at least within BOSS clustering uncertainties on the scales considered in this paper.} However, there are some interesting $1\sigma$-level shifts between the two fits that could become statistically significant in future studies with higher signal-to-noise measurements. 

\red{Besides improving the measurement precision, we can also untangle potential degeneracies between cosmologies and secondary biases in galaxy--halo connection models by incorporating additional summary statistics. Galaxy--galaxy lensing is directly informative of the galaxy--matter cross-correlation, which linearly depends on galaxy bias while the galaxy--galaxy auto-correlation function depends on galaxy bias squared. Thus, the addition of galaxy--galaxy lensing could effectively decouple the galaxy secondary bias models from dark matter clustering.} Other informative small-scale statistics beyond the galaxy auto-correlation functions include the galaxy 3-point correlation function and its various compressions \citep[e.g.][]{2017Yuan, 2015bGuo, 2015Slepian, 2011McBride, 2004Wang}, marked correlation functions\citep[e.g.][]{2021Massara, 2019Satpathy, 2016White, 2006Skibba}, and density-based statistics such as the $k$-th nearest neighbor statistics \citep[kNN, e.g.][]{2021Wang, 2021Banerjee, 2021bBanerjee}, counts-in-cell statistics \citep[e.g.][]{2020Repp, 2019Wang, 2011Yang, 2009Reid}, and void probability functions \citep[e.g.][]{2019Walsh, 2009Betancort, 2008Tinker, 1987Maurogordato}.  

\subsection{HOD constraints}

\begin{figure*}
    \centering
    \hspace*{-0.3cm}
    \includegraphics[width = 7.2in]{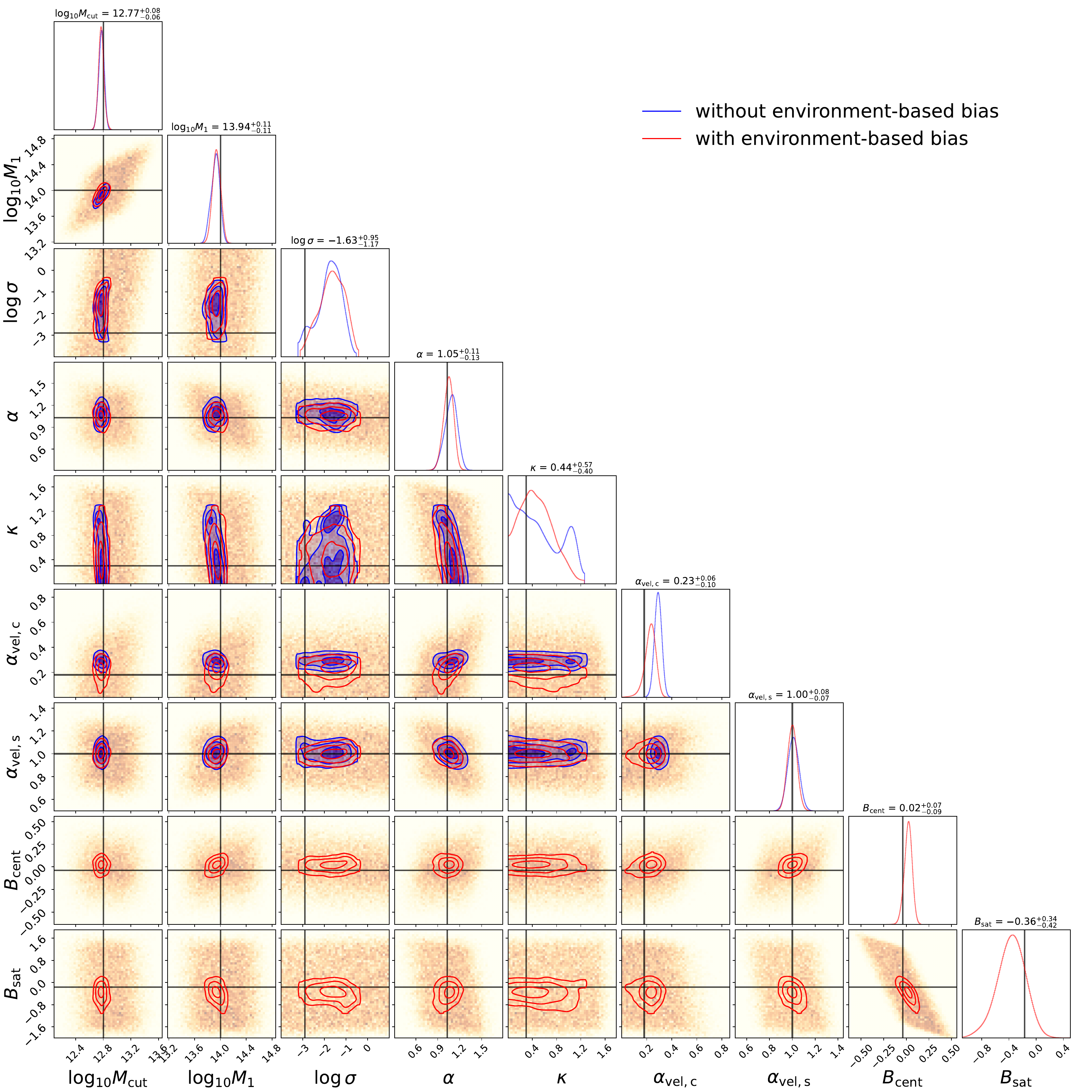}
    \vspace{-0.3cm}
    \caption{The 2D marginalized constraints on HOD parameters from the CMASS small-scale RSD measurement. The red and blue contours showcase the constraints with and without enabling environment-based secondary biases in the HOD, respectively. The contours represent the $1,2,3\sigma$ constraints. The underlying heatmaps show the prior volume. The titles on the 1D constraints show the posterior mean and $2\sigma$ constraints in the with environment-based bias case, ($95\%$ C.L.). The black lines denote the best-fit values at fixed fiducial cosmology, taken from \citet{2021bYuan}.}
    \label{fig:corner_hod}
\end{figure*}

Figure~\ref{fig:corner_hod} showcases our marginalized HOD constraints, with and without environment-based biases. Again, the underlying heatmap shows the marginalized prior volume. The black lines represent the best-fit HOD values at Planck cosmology, quoted from \citet{2021bYuan}. We see that we obtain strong constraints on most HOD parameters, as expected for a small-scale full-shape clustering analysis. The only parameter that does not appear well constrained by the data is $\kappa$, which together with $M_\mathrm{cut}$, determines the minimum halo mass to host a satellite $M_\mathrm{min} = \kappa M_\mathrm{cut}$. However, satellite predominantly occupy the highest mass halos and its occupation is not strongly affected by the minimum mass cut-off, resulting in $\kappa$ having a very weak signature on clustering. 

We obtain strong constraints on the halo mass parameters, $\log M_\mathrm{cut}$ and $\log M_1$, which partially determine the typical halo mass of central and satellite galaxies, respectively. The best-fit values are also consistent with the fixed cosmology values shown in the black lines. $\log \sigma$ and $\alpha$, which control the shape of the central and satellite occupation functions, respectively, are also reasonably well constrained. The constraints on $\alpha$ are consistent with that of \citet{2021bYuan}, whereas the constraints on $\sigma$ is higher compared to \citet{2021bYuan}, suggesting a slower turn-off in central occupation at lower halo masses. Comparing the red and blue contours, the environment-based biases do not appear to bias or dilute these constraints. 

From the cosmology and HOD constraints, we can also derive the satellite fractions and average halo mass of galaxies. With environment-based bias, the inferred satellite fraction is $f_\mathrm{sat} = 0.126\pm 0.010$, and the average host halo mass is $\langle\log M\rangle = 13.36\pm 0.03$, where $M$ assumes the unit of $h^{-1}M_\odot$. Without environment-based bias, the inferred satellite fraction is $f_\mathrm{sat} = 0.122\pm 0.006$, and the average host halo mass is $\langle\log M\rangle = 13.38\pm 0.03$. We see that the satellite fraction and the average halo mass are both very well constrained and robust against the inclusion of environment-based biases. These inferred values are also consistent with previous fits for CMASS LRGs \citep[e.g.][]{2015aGuo}. 

Moving on to the velocity bias parameters, we find a somewhat different best-fit central velocity bias with and without environment-based bias. \red{Without environment-based bias, a higher $\alpha_\mathrm{vel, c}= 0.29\pm0.03$ is preferred, whereas a lower value of $\alpha_\mathrm{vel, c}= 0.23\pm 0.04$ is preferred with environment-based bias, a roughly $2\sigma$ difference. This potentially points to a mild degeneracy between environment-based bias and velocity bias. By allowing galaxies to preferentially occupy halos in denser environments, these galaxies would occupy deeper potential wells and thus have higher peculiar velocities, reducing the need to invoke additional velocity bias. Comparing to the fixed-cosmology constraints shown in black, our new central velocity bias is biased slightly higher. This is likely due to the lower inferred $\sigma_8$ in this analysis, which corresponds to a lower peculiar velocities of the halos. Thus, larger velocity bias is required to reproduce the observed FoG. This degeneracy is also shown in the 2D posteriors in Figure~\ref{fig:corner_full}.} For satellites, we do not see any evidence for velocity bias, in both cases. This is consistent with  previous RSD analyses at fixed cosmology \citep[e.g.]{2021bYuan, 2015aGuo}, and HOD analysis of LRG-like samples in hydrodynamical simulations \citep{2022Yuan}. 

\red{For environment-based biases, we do not find evidence for a non-zero $B_\mathrm{cent}$, but we do find a negative best-fit $B_\mathrm{sat}$ at the $2\sigma$ level. This suggests that the satellites preferentially occupy halos in denser environments at fixed halo mass. This is statistically consistent with our findings at fixed cosmology. The decrease in constraining power on $B_\mathrm{cent}$ and $B_\mathrm{sat}$ compared to \citet{2021bYuan} also possibly reflects a partial degeneracy between environment-based bias and cosmological parameters.} Our constraints on environment-based bias are also supported by our findings in hydrodynamical simulations, where we find no environment-based bias for LRG centrals but strong environment-based bias for LRG satellites \citep{2022Yuan}. This result is also significant because it is the first environment-based bias constraints marginalized over cosmology. The consistency between independent observational constraints and simulation constraints also combine to make a compelling case for the existence of environment-based secondary bias.

\section{Discussion}
\label{sec:discussion}
\subsection{Comparison to existing $f\sigma_8$ constraints}
\label{subsec:compare_sigma8}
\begin{figure*}
    \centering
    \hspace*{-0.3cm}
    \includegraphics[width = 6in]{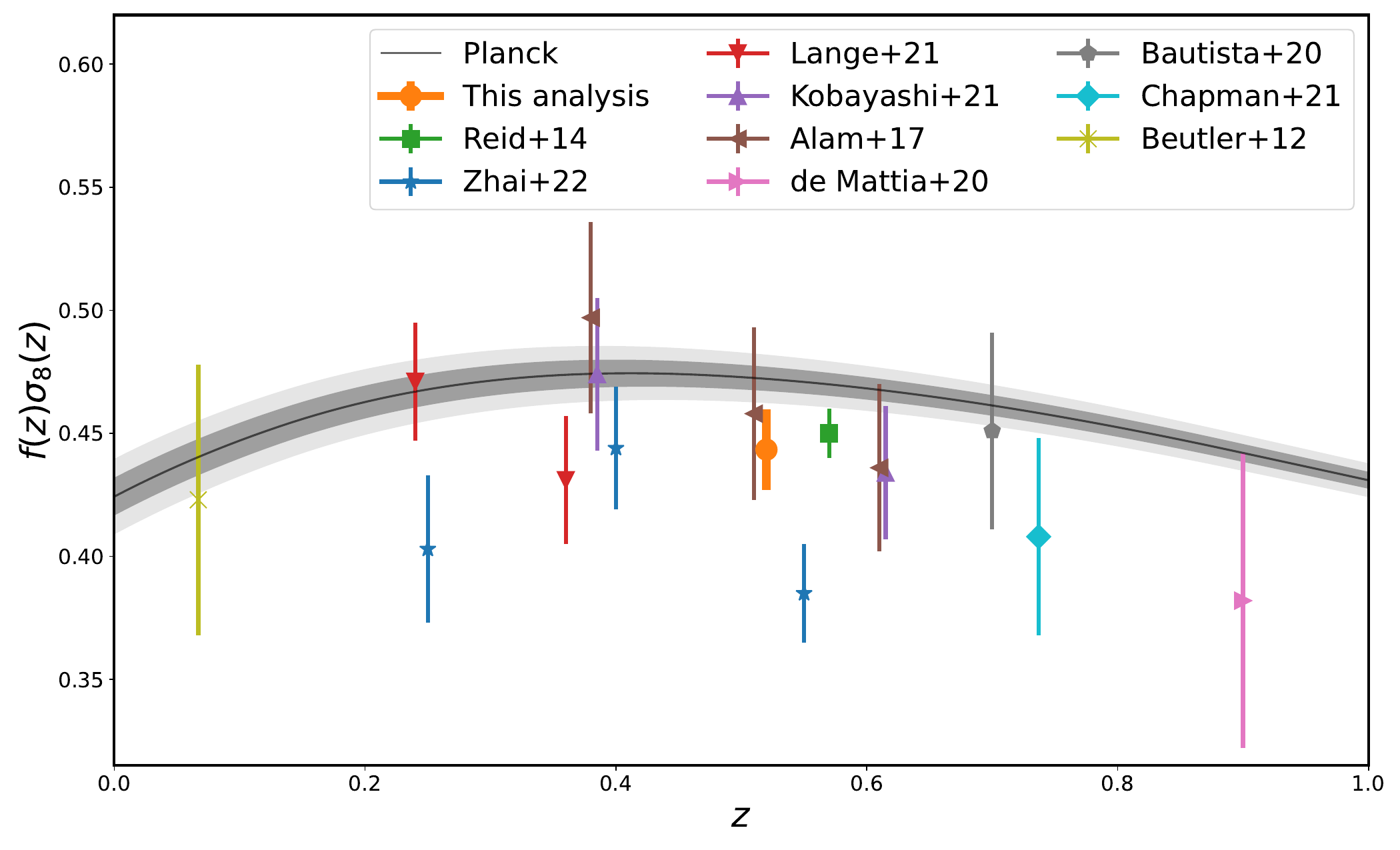}
    \vspace{-0.3cm}
    \caption{The marginalized constraints on the structure growth rate $f\sigma_8(z)$ of our fiducial analysis (including environment-based bias) compared to other clustering constraints. We show Planck CMB constraints in black (the different shades correspond to $1\sigma$ and $2\sigma$). Additionally, we show clustering constraints from BOSS CMASS small-scale RSD \citep{2014Reid}, BOSS LOWZ small-scale RSD \citep{2021Lange}, BOSS full-shape power spectrum \citep{2021Kobayashi}, BOSS large-scale RSD+BAO \citep{2017Alam}, eBOSS small-scale RSD \citep{2021Chapman}, eBOSS large-scale RSD+BAO \citep{2021Bautista, 2021deMattia}, and the 6dF Galaxy Survey \citep{2012Beutler}. Our constraints are tighter than most other analyses, resulting in a 2-3$\sigma$ tension with Planck. The constraints of \citet{2014Reid} are overly narrow due to over-simplifications in their velocity re-scaling method. }
    \label{fig:fsigma8}
\end{figure*}

Figure~\ref{fig:fsigma8} compares our growth rate of structure constraints $f\sigma_8(z)$ with previous clustering analyses using BOSS and extended BOSS data \citep[eBOSS;][]{2016Dawson}. \red{To obtain $f\sigma_8$ constraints at an arbitrary redshift $z$, we use \textsc{Colossus}, which allows us to analytic evolve redshift-dependent cosmology parameters to arbitrary redshifts. Specifically, we draw a uniform subsample from our cosmology posterior chain and use \textsc{Colossus} to calculate the $f\sigma_8(z)$ for each point in the subsample.} For our fiducial analysis including environment-based bias, we obtain 
\begin{equation}
    f\sigma_8(z_\mathrm{eff}) = 0.444_{-0.016}^{+0.016}  \ \ (68\%,\  \mathrm{with\ environment\text{-}based\ bias}),
    \label{equ:fsig8_z}
\end{equation}
where the effective redshift of the sample $z_\mathrm{eff} = 0.52$. Without environment-based bias, the $1\sigma$ uncertainty further reduces to $\pm 0.012$. 

We showcase the Planck 1--2$\sigma$ constraints with the black bands. The other late-time growth rate constraints can be further grouped into large-scale analyses and small-scale analyses, with $30h^{-1}$Mpc as a rough delinearization between the two regimes. The large-scale results rely on analytical models of large-scale features such as the BAO and RSD and are thus  less sensitive to non-linear evolution and the details of galaxy--halo connection modeling. Such studies include \citet{2017Alam} using BOSS large-scale RSD+BAO, \citet{2021Bautista} and \citet{2021deMattia} using eBOSS large-scale RSD+BAO, and \citet{2012Beutler}, which uses the 6dF Galaxy Survey at lower redshift.

The second category of small-scale analyses extract cosmological information from the non-linear clustering at below roughly $30h^{-1}$Mpc by leveraging cosmological simulations and more sophisticated galaxy--halo connection modeling. \citet{2014Reid} represents a first attempt at constraining $f\sigma_8$ with CMASS small-scale RSD by leveraging a single dark matter only simulation. The authors employed a velocity re-scaling technique, relying on the assumption that changes in $f\sigma_8$ are completely degenerate with a simple linear scaling of the velocity field on all scales. However, \citet{2019Zhai} showed that this is an over-simplification and results in artificially tight constraints on $f\sigma_8$. 

Later small-scale studies adopted emulator approaches relying on suites of cosmological simulations and a flexible interpolation scheme. \citet{2021Kobayashi} applied this approach to BOSS full-shape galaxy power spectrum, deriving slightly stronger constraints than large-scale studies. However, their analysis uses a simple vanilla HOD excluding potential secondary biases. \citet{2021Lange} applied the evidence modeling approach, which emulates the cosmology dependence of the likelihood function after marginalizing over the HOD. The authors analyzed the small-scale RSD of a small sample of BOSS LOWZ LRGs and found constraints competitive with other small-scale analyses, while including a concentration-based prescription of secondary bias. \citet{2022Zhai} and \citet{2021Chapman} respectively analyzed the small-scale clusteirng of BOSS and eBOSS LRGs with an emulator built on the \textsc{Aemulus} simulation suite \citep{2019DeRose}.

Figure~\ref{fig:fsigma8} shows that this analysis adds to growing evidence of a potential $f\sigma_8$ (or $S_8$) tension, with several measurements of late-time galaxy clustering and lensing resulting in constraints on $f\sigma_8$ and $S_8$ that are systematically lower than Planck CMB constraints. Our $f\sigma_8$ posterior mean is closer to Planck constraints percentage-wise compared to many other studies, but still at roughly $2\sigma$ tension due to the smaller error bars. Comparing the constraints from small-scale analyses to those from large-scale analyses, one can clearly see the potential gain in cosmological information when leveraging non-linear clustering. However, there is still some disagreement amongst the small-scale analyses despite a qualitative agreement in favoring a low $f\sigma_8$ relative to Planck. Our results are statistically consistent with all other small-scale analyses except for the highest redshift bin of \citet{2022Zhai}, which finds a particularly low $f\sigma_8$ that is in $\sim2.9\sigma$ tension with our results. A full exploration of this disagreement is beyond the scope of this paper, but is necessary to ensure small-scale simulation-based analyses account for all the necessary systematics. Some key systematic differences between the two analyses include the simulation suite and halo finder used, the range of scales probed, and the range of cosmological models explored. Galaxy--halo connection modeling represents another potential source of systematics. However, both analyses use similar HOD models extended with velocity bias and environment-based bias. Our specific choice of HOD model is informed on hydrodynamical simulations. We also test the effect of a variety of physically motivated model extensions (section~\ref{subsec:additional_extensions}) and find our $f\sigma_8$ constraints to be robust against such extensions. 

Compared to other small-scale analyses, we obtain somewhat tighter growth rate constraints. This is likely due to a combination of several factors. \red{First of all, the \textsc{AbacusSummit} simulations have large volume and an extensive cosmology grid. The large volume helps reduce sample variance, as shown in section~\ref{subsec:cov_bias}. The cosmology grid is the largest of its kind and contains additional sampling along parameter directions such as $\sigma_8$ and $\omega_\mathrm{cdm}$. The cosmologies are also sampled along well-constrained quantities such as the acoustic peak scale. These design choices should help increase density of training sample in the parameter space of interest and reduce the resulting emulator error. The sample density of HOD training set is also increased via the hybrid MC + emulator approach. Existing HOD emulators aim to accurately predict observables over an overly expansive parameter space by spanning the parameter space with a latin hypercube. However, such a training set covers many regions that are ruled out by the data. With our hybrid approach, we construct the HOD training sets only within realistic regions of the parameter space, thus concentrating the limited computational power (limited by memory) and model flexibility on improving accuracy in a smaller region of the parameter space. The combination of these different design choices should minimize emulator errors and thus reduce the uncertainties in our posterior constraints.}

\red{
Finally, we caution that our estimation of the emulator errors is rather approximate and averaged over only a few test cosmologies. In principle, the emulator error is a function of cosmology and HOD. We showed in section~\ref{subsec:emu_test} that the typical emulator error varies from as low as $<10\%$ of the data error at Planck cosmology to close to $50\%$ of data error at other cosmologies. This suggests there is uncertainty in the final error bar we quote for $f\sigma_8$. However, our best-fit cosmology is indeed close to the center of the cosmology grid and within the range probed by the test cosmologies. As a result, the emulator error we reported is likely a reasonable estimate and that the final error bars are still dominated by data error and not by emulator error. 
}

\subsection{Effects of HOD extensions}
\label{subsec:additional_extensions}
\begin{figure}
    \centering
    \hspace*{-0.3cm}
    \includegraphics[width = 3.2in]{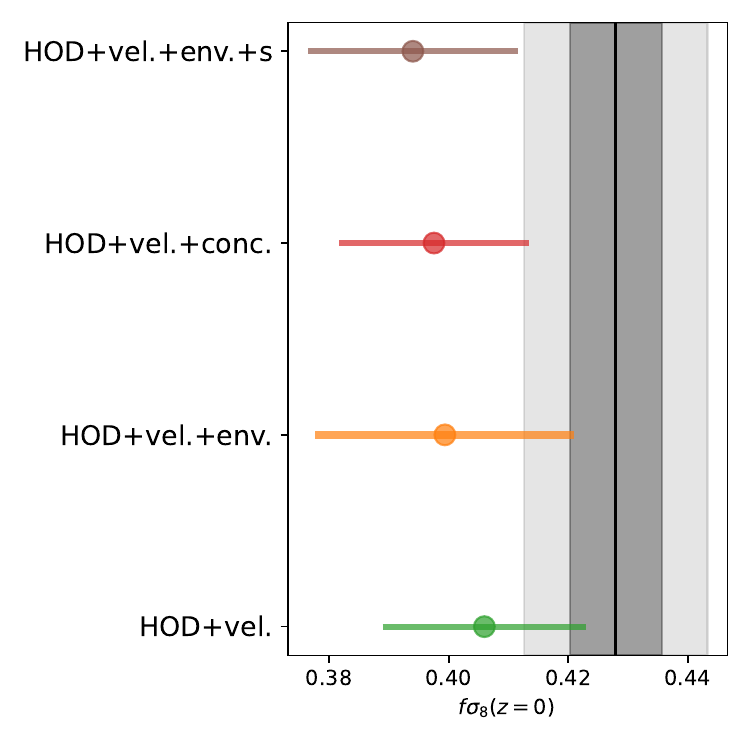}
    \vspace{-0.3cm}
    \caption{The marginalized constraints on the structure growth rate $f\sigma_8(z = 0)$ for different HOD model choices. The fiducial choice of a vanilla HOD extended with velocity bias and environment-based bias is shown in orange. The green line shows the baseline scenario that includes only the vanilla HOD and velocity bias. The red line corresponds to the vanilla HOD plus velocity bias plus concentration-based secondary bias, the commonly assumed assembly bias model. The brown line corresponds to the fiducial model plus a satellite radial distribution parameter ``s''. The black line and shaded regions correspond to Planck posterior mean and $1-2\sigma$ constraints. We see that while the mean inferred $f\sigma_8$ is consistent across different HOD model choices, the posterior spread is somewhat dependent on model choices. 
    }
    \label{fig:fsigma8_hod}
\end{figure}
While HOD modeling enables us the flexibility to explore clustering on smaller scales, it also allows for arbitrary model choices that can potentially bias and dilute the cosmology constraints. Fortunately, the \ahod\ framework is designed to accommodate a variety of HOD extensions while maintaining high computational efficiency. In this section, we test how different HOD model choices affect the constraints on growth rate $f\sigma_8(z = 0)$.

Figure~\ref{fig:fsigma8_hod} showcases the $f\sigma_8(z = 0)$ $1\sigma$ constraints for four different HOD choices. Starting from the bottom, the green marker denotes a baseline model that consists of the vanilla 5-parameter HOD model plus velocity bias. The orange marker corresponds to the fiducial model that also includes the environment-based secondary biases in addition to the baseline model. The red marker corresponds to the baseline model plus a concentration-based secondary model, which is the commonly assumed assembly bias parameterization. The brown marker on top represents the fiducial model shown in orange plus a radial distribution parameter ``s'' that allows the satellite distribution to deviate from the halo density PDF. The vertical black line and shaded regions denote the Planck 2018 $1$ and $2\sigma$ constraints. 

We see that the posterior mean of $f\sigma_8$ is consistent across these four different HOD model choices. This suggests that our best-fit cosmology is robust against HOD models. The consensus growth rate values are all 7-8$\%$ lower than Planck, suggesting a mild tension between small-scale RSD measurements at late times and the CMB. Finally, the constraining power on $f\sigma_8$ is somewhat dependent on HOD modeling choices. Particularly, the inclusion of environment-based secondary biases seems to enlarge the constraints, probably due to a degeneracy between environment-based bias and growth rate of structure. However, it is important to note that the inclusion of secondary biases do not remove the mild tension in growth rate. 

The consistent cosmology inference between different HOD choices is an important result as it demonstrates that we can obtain robust and tight constraints on cosmology from galaxy clustering on small scales. This conclusion is particularly encouraging in the context of preparing for DESI. Applying our inference pipeline to DESI small-scale redshift-space clustering measurements should greatly compliment the traditional large-scale BAO and RSD analyses and contribute significantly to the amount of extracted cosmological information. 

However, our tests are limited to a single HOD framework and a single set of simulations. We plan on conducting additional mock challenge tests where we recover the underlying cosmology from mock RSD measurements generated on different simulation suites and with other types of galaxy--halo connection models, specifically those that do not rely on an HOD model but instead leverages semi-analytic models or hydrodynamical simulations. This is an ongoing effort, and we reserve the corresponding analysis for a future paper. 

\subsection{galaxy--galaxy lensing prediction}

A separate well known tension exists between galaxy clustering and galaxy--galaxy lensing (g--g lensing), known as ``lensing is low''. \citet{2017Leauthaud} first identified a discrepancy of 20-40$\%$ between their measurements of g--g lensing for CMASS galaxies and a mock prediction generated at Planck cosmology that match the CMASS projected correlation function \citep[][see Figure~7 of \citealt{2017Leauthaud}]{2014Reid, 2016Saito}. \citet{2019Lange, 2020Lange} extended this result by finding a similar ${\sim} 25\%$ discrepancy between the projected clustering measurement and the g--g lensing measurement in the BOSS LOWZ sample. In \citet{2019Yuan}, we reaffirmed this tension by fitting simultaneously the projected galaxy clustering and g--g lensing with an extended the HOD incorporating a concentration-based assembly bias prescription, at fixed cosmology. 
In \citet{2021Yuan, 2021bYuan}, we found that the inclusion of an environment-based secondary bias can significantly reduce ($\sim 10\%$) the predicted lensing signal when constrained on the redshift-space 2PCF at fixed cosmology. However, the inclusion of environment-based bias alone is not sufficient in reconciling the ``lensing is low'' tension. In this analysis, we extend the RSD analysis from Planck cosmology to variable cosmology, and we can similarly make g--g lensing predictions marginalized over both the extended HOD and cosmology posteriors. 

\begin{figure}
    \centering
    \hspace*{-0.3cm}
    \includegraphics[width = 3.5in]{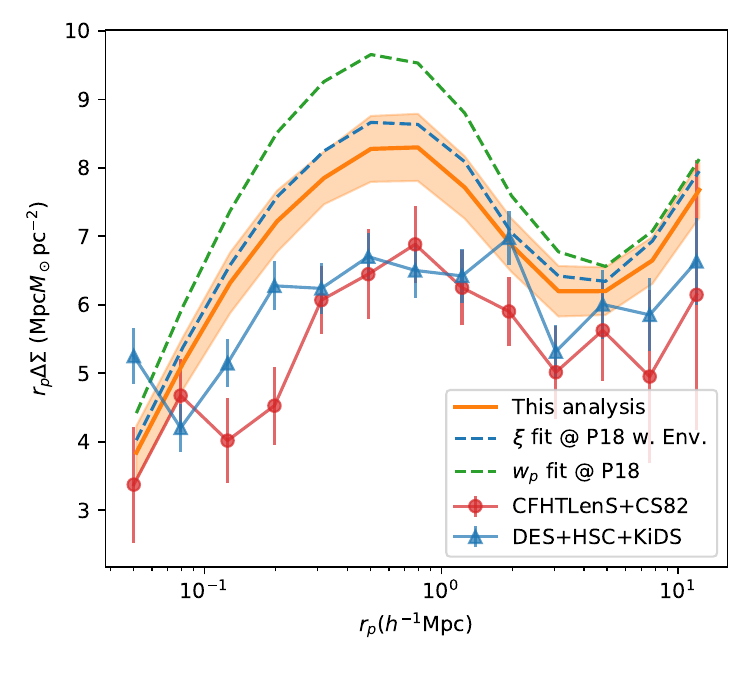}
    \vspace{-0.3cm}
    \caption{The marginalized prediction of galaxy--galaxy lensing of various clustering analyses compared to data. The data measurements are shown in the red and blue curves, where the red shows CMASS lenses against CFHTLenS+CS82 source galaxies and the blue shows a more updated measurement of the same lenses against DES+HSC+KiDS sources. The green dashed line shows the naive prediction using a vanilla 5-parameter HOD constrained only on the projected correlation function $w_p$ at fixed cosmology, in clear tension with the observations. The blue dashed line represents the prediction from the small-scale RSD analysis at Planck cosmology conducted in \citet{2021bYuan}, which included the effects of environment-based bias but not of cosmology. The orange line shows the mean posterior prediction of this fiducial analysis, where both the effects of environment-based bias and cosmology are accounted for. The shaded region shows the $1\sigma$ uncertainties, constituting mostly of the cosmology+HOD posterior constraints and the lensing emulator error. Comparing the orange curve to the green curve, we conclude that the combined effect of environment-based bias and lower growth rate accounts for approximately $50\%$ of the ``lensing is low'' tension.}
    \label{fig:ggl}
\end{figure}

To derive the lensing prediction, we train a lensing emulator following the same procedure as we trained the RSD emulator. We use the same cosmology and HOD training sets. We then use the trained lensing emulator to predict g--g lensing over the cosmology+HOD posterior chains that were constrained on the small-scale RSD measurement. We summarize the marginalized lensing predictions in Figure~\ref{fig:ggl}. The red and blue lines show two independent lensing measurements of the CMASS sample, with the red using the Canada France Hawaii Telescope Lensing Survey \citep[CFHTLenS,][]{2012Heymans, 2013Miller} and the Canada France Hawaii Telescope Stripe 82 Survey \citep[CS82,][]{2013Erben} as the source samples, and the blue using the more recent photometric samples from DES Y1 \citep[][]{2018Drlica, 2018Zuntz, 2016DES}, HSC \citep[][]{2018Aihara, 2019Aihara, 2018Miyazaki}, and KiDS datasets \citep[][]{2019Kuijken, 2021Giblin}. The blue curve is generated using the publicly available \textsc{dsigma} code (Lange et al. in prep)\footnote{\url{https://github.com/johannesulf/dsigma}}. The orange curve and shaded area showcase the marginalized lensing predictions from our fiducial analysis assuming an HOD model extended with velocity bias and environment-based secondary bias. The shaded area corresponds to 1$\sigma$ posterior uncertainties. The dashed blue curve represents the best-fit prediction from \citet{2021bYuan}, where we fit the same CMASS small-scale RSD measurement with the same fiducial extended HOD model but at fixed Planck 2018 cosmology. Thus, comparing the blue dashed line and the orange line, we see that the combined effect of a lower inferred growth rate and environment-based bias pushes the predicted lensing amplitude further towards agreement with observations. For reference, the green dashed line represents the naive lensing prediction with a vanilla 5-parameter HOD model constrained against the projected galaxy correlation function $w_p$ at Planck cosmology (reproducing the solid red curve in Figure~11 of \citealt{2021bYuan}). Comparing the orange line and the green dashed line, we see that a more robust prediction utilizing the full-shape clustering information and accounting for both environment-based secondary bias and variable cosmology results in a $15\%$ reduction to the predicted lensing signal, which accounts for approximately $50\%$ of the lensing is low tension. 
Other potential sources of tension include baryonic effects \citep{2020Amodeo}, and possibly additional unaccounted for observational systematics \citep[][]{2022Amon}. We speculate that a future joint analysis of galaxy clustering and galaxy--galaxy lensing that accounts for environment-based bias, variable cosmology, and baryonic effects can potentially full resolve the ``lensing is low'' tension. 

We also make marginalized lensing predictions for the baseline model that does not include any secondary biases, and the model including a concentration-based secondary bias. Both of these model prescriptions result in lensing amplitudes approximately a few percent larger the fiducial model with environment-based bias. We do not showcase these cases for brevity. 

\subsection{DESI outlook}
\label{subsec:desi}
While the CMASS effective volume is approximately $5$Gpc$^3$, the eventual DESI LRG sample is expected to span an effective volume of 40Gpc$^3$ \citep[e.g.][]{2019Chuang}, representing roughly an order of magnitude increase. This translates to an expected sample variance with DESI LRGs that is approximately 1/8 that of BOSS CMASS. Within the regime where the primary source of uncertainty is sample variance in the data, the expected errorbar on $f\sigma_8$ should tighten by approximately a factor of 3. The challenge for small-scale emulator analysis such as this is to ensure that the emulator error is subdominant to that of sample variance. 

Referring back to Figure~\ref{fig:test_emulator_rel}, we see that our current emulator error is typically 30--40$\%$ that of the CMASS sample variance, or 10--20$\%$ in terms of covariance matrix. This means that at projected DESI LRG effective volume, our emulator error would be no longer sub-dominant but roughly comparable to DESI sample variance. However, for the initial DESI cosmology analysis, which we expect to only utilize approximately $20\%$ of the eventual DESI volume, our current emulator scheme should offer plenty of precision. 

To utilize the full DESI LRG volume, it is important to explore ways of further improving the emulator errors. One potential avenue of achieving smaller errors is in implementing more optimal surrogate models. It is possible that Gaussian Process is not the optimal model for this problem and a more regularized model such as polynomial expansion could further suppress emulator errors. However, the main limitation of our framework is the sparse cosmology sampling. While there exists additional N-body simulation suites that provide additional cosmology coverage, but they utilize different N-body codes and halo finders, making HOD parametrizations hard to reconcile. 
Instead, we propose an iterative approach where we plan on running additional \textsc{Abacus} boxes/cosmologies in high-likelihood regions of cosmology space. Operationally, this means we will use the existing \textsc{AbacusSummit} for the first round of inference with DESI data. Then we sample additional cosmologies in the favored regions of cosmology parameter space. Then we re-train our models on the expanded cosmology grid and re-run the inference to achieve stronger constraints. We can iterate this step until a desired emulator accuracy is achieved. 

Another significant opportunity with DESI lies in accurate measurements of small-scale clustering for galaxy tracers other than LRGs, providing yet another important lever arm on cosmology. For example, we expect the effective volume of the DESI emission line galaxy (ELG) sample to be equal or larger than that the LRGs due to an extensive redshift coverage and higher number densities. It remains to be seen whether our hybrid framework will yield emulator errors competitive with DESI ELG clustering measurements. We reserve a detailed hybrid emulator analysis of ELGs for a future paper.


\section{Conclusions}
\label{sec:conclude}

In this paper, we have devised a novel hybrid MCMC-emulator inference framework for small-scale galaxy clustering that is sufficiently precise to leverage the DESI Y1 LRG sample. We applied the framework on CMASS data and found stringent constraints on $f\sigma_8$. Our constraints are at mild (2--3$\sigma$) tension with Planck, but consistent with other late-time large-scale structure constraints. These constraints should tighten considerably when DESI data become available, finally elucidating any potential tension with Planck. We summarize the key takeaways as follows:
\begin{itemize}
  \item Our approach successfully suppresses emulator errors through larger simulations, denser cosmology+HOD sampling, and data-informed training. As a result, our analysis of CMASS achieves the strongest constraints on $f\sigma_8$ compared to other galaxy clustering studies (Figure~\ref{fig:fsigma8}). We anticipate that our approach offers sufficient accuracy to take advantage of the effective volume of DESI Y1 LRG sample, and possibly the ELG sample. 
  
  \item Our growth rate constraints using CMASS data are $f\sigma_8(z=0) = 0.407^{+0.018}_{-0.023}$ when not including environment-based secondary bias, and $f\sigma_8(z=0) = 0.398^{+0.021}_{-0.018}$ when including environment-based bias. These results are in $3\sigma$ and $2.3\sigma$ tension with Planck CMB constraints, respectively. We expect these errorbars to shrink significantly with DESI data. 
  
  \item We demonstrate that our $f\sigma_8$ constraint is robust against HOD model choices (Figure~\ref{fig:fsigma8_hod}), serving as a important test of the cosmological constraining power of small-scale clustering. This also shows that HOD details such as secondary biases can not explain the emerging $\sigma_8$ tension. As far as we are aware, this robustness test is the first of its kind. We plan on conducting additional mock challenges to test robustness against different simulation suites and non-HOD galaxy--halo connection models. 
   
  \item We continue to find strong evidence for the inclusion of environment-based secondary bias in the HOD model. Specifically, we find that satellite LRGs preferentially occupy halos in denser environments, consistent with previous RSD analysis and HOD studies using hydrodynamical simulations. 

  \item We find that the lower $f\sigma_8$ and the environment-based bias combine to give a $15\%$ reduction in the predicted galaxy--galaxy lensing signal, contributing approximately $50\%$ to the ``lensing is low'' tension.
\end{itemize}

Ultimately, this analysis framework is designed for DESI, where we fully expect to obtain significantly stronger cosmology+HOD constraints. DESI should also offer sufficient redshift coverage for a tomographical analysis that examines the potential redshift evolution of the growth rate. Another important clustering science opportunity with DESI lies in multi-tracer analyses, where the additional volume available in other galaxy types and the additional statistics in their cross-correlations should further strength the constraining power of the small-scale lever arm.



\section*{Acknowledgements}
We would like to thank Sean McLaughlin, Johannes Lange, Nick Kokron, Alexie Leauthaud, Joe DeRose, Andrew Hearin, Josh Speagle, and others for useful feedback and suggestions in various stages of this analysis.

This work was supported by U.S. Department of Energy through grant DE-SC0013718 and 
under DE-AC02-76SF00515 to SLAC National Accelerator Laboratory, NASA ROSES grant 12-EUCLID12-0004, NSF PHY-2019786, and the Simons Foundation.
SB is supported by the UK Research and Innovation (UKRI) Future Leaders Fellowship [grant number MR/V023381/1].

This work used resources of the National Energy Research Scientific Computing Center (NERSC), a U.S. Department of Energy Office of Science User Facility located at Lawrence Berkeley National Laboratory, operated under Contract No. DE-AC02-05CH11231.
The {\sc AbacusSummit} simulations were conducted at the Oak Ridge Leadership Computing Facility, which is a DOE Office of Science User Facility supported under Contract DE-AC05-00OR22725, through support from projects AST135 and AST145, the latter through the Department of Energy ALCC program.

Some of the computing for this project was performed on the Sherlock cluster. We would like to thank Stanford University and the Stanford Research Computing Center for providing computational resources and support that contributed to these research results.

This project used data from SDSS-III BOSS. Funding for SDSS-III has been provided by the Alfred P. Sloan Foundation, the Participating Institutions, the National Science Foundation, and the U.S. Department of Energy Office of Science. The SDSS-III web site is http://www.sdss3.org/.

SDSS-III is managed by the Astrophysical Research Consortium for the Participating Institutions of the SDSS-III Collaboration including the University of Arizona, the Brazilian Participation Group, Brookhaven National Laboratory, Carnegie Mellon University, University of Florida, the French Participation Group, the German Participation Group, Harvard University, the Instituto de Astrofisica de Canarias, the Michigan State/Notre Dame/JINA Participation Group, Johns Hopkins University, Lawrence Berkeley National Laboratory, Max Planck Institute for Astrophysics, Max Planck Institute for Extraterrestrial Physics, New Mexico State University, New York University, Ohio State University, Pennsylvania State University, University of Portsmouth, Princeton University, the Spanish Participation Group, University of Tokyo, University of Utah, Vanderbilt University, University of Virginia, University of Washington, and Yale University.

\section*{Data Availability}

The simulation data are available at \url{https://abacussummit.readthedocs.io/en/latest/}. The \ahod\ code package is publicly available as a part of the \textsc{abacusutils} package at \url{http://https://github.com/abacusorg/abacusutils}. Example usage can be found at \url{https://abacusutils.readthedocs.io/en/latest/hod.html}.



\bibliographystyle{mnras}
\bibliography{biblio} 

\appendix

\section{Mock covariance matrix}
\label{sec:cov_comapre}
For our CMASS jackknife covariance matrix $\boldsymbol{C}_\mathrm{data}$ calculation, we chose the number of jackknife regions to be sufficiently large to yield a well-conditioned covariance matrix, but each region has a rather small volume as a result, raising the concern of this process might have introduced additional sample variance into the covariance matrix. Here we validate that this jackknife covariance matrix is consistent with a mock-based covariance matrix computed on a much larger volume. 

To compute a mock-based covariance matrix, we take a fiducial HOD fit to the target data vector, the CMASS $\xi(r_p, \pi)$, at our fiducial Planck cosmology, and apply it to the 25 random phase boxes. We then segment each box into 8 smaller $1h^{-1}$Gpc$^3$ boxes to arrive at 200 sub-boxes, each comparable to the CMASS sample volume, which is approximately 1.7$h^{-1}$Gpc$^3$. Finally, we compute the mock-based covariance by taking the variance from the 200 sub-boxes and then normalize to the CMASS volume. In total, the mock-based covariance matrix is computed from a volume 118 times that of our CMASS sample. 

\begin{figure}
    \centering
    \hspace*{-0.3cm}
    \includegraphics[width = 3.7in]{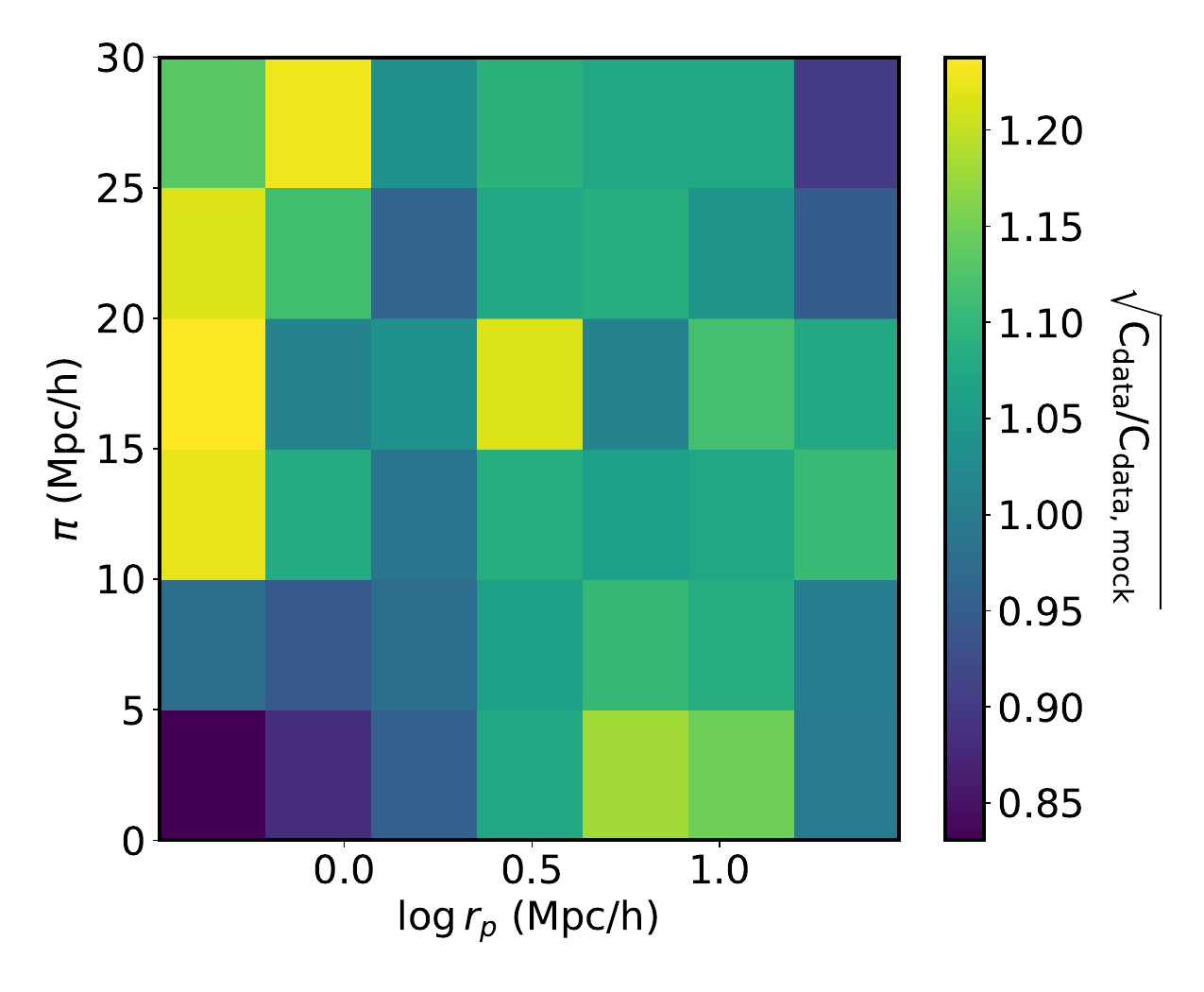}
    \vspace{-0.3cm}
    \caption{The square root ratio of the CMASS jackknife covariance matrix diagonal to the mock-based covariance matrix diagonal. This shows that the two covariance matrices have consistent amplitudes along the diagonal.}
    \label{fig:cov_compare_diag}
\end{figure}

\begin{figure}
    \centering
    \hspace*{-0.3cm}
    \includegraphics[width = 3.7in]{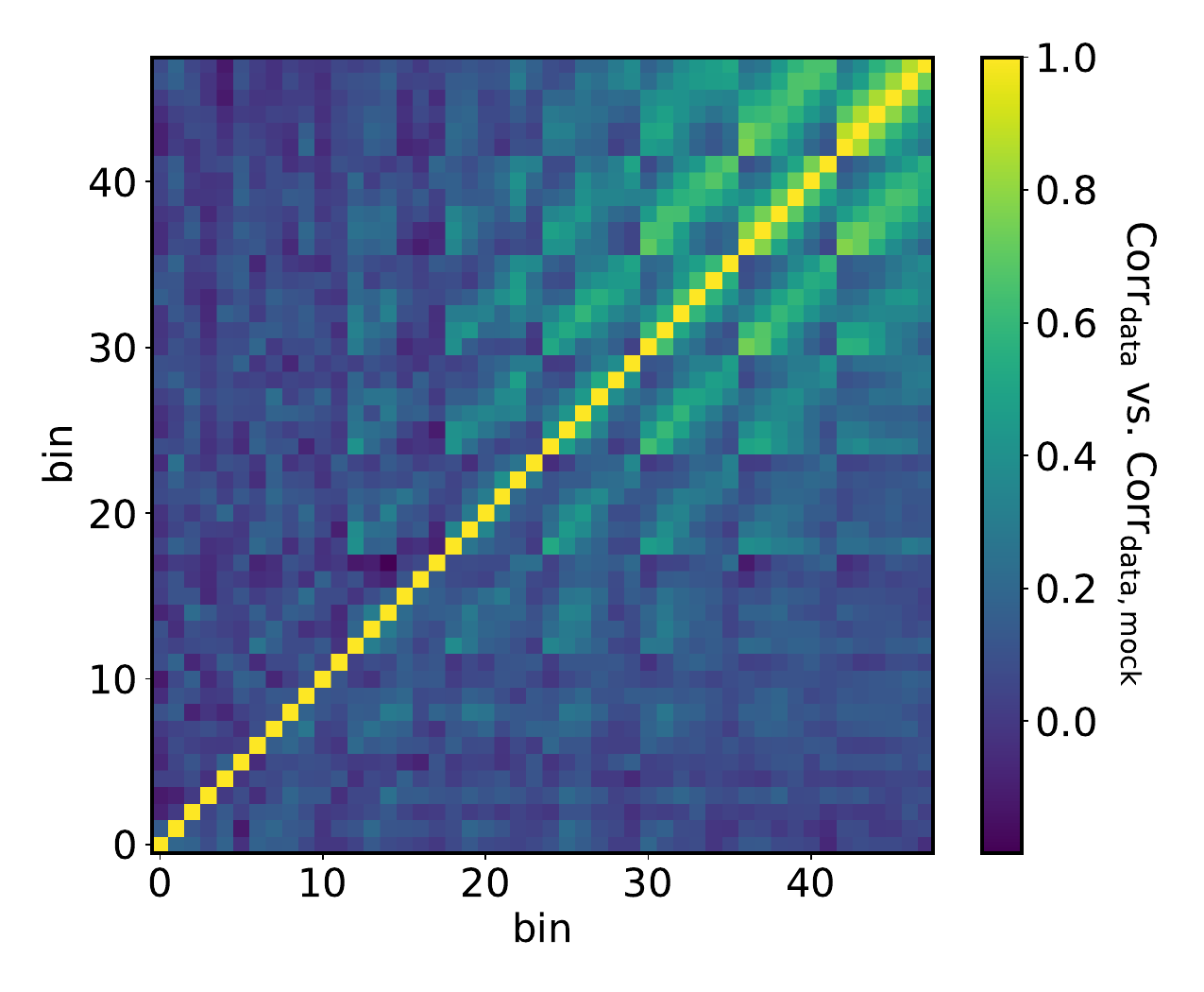}
    \vspace{-0.3cm}
    \caption{The correlation matrix of the jackknife covariance matrix in the top left triangle and the correlation matrix of the mock-based covariance matrix in the lower right triangle. This shows that the two covariance matrices have consistent off-diagonal structure.}
    \label{fig:cov_compare_corr}
\end{figure}

To visualize the comparison of the two covariance matrices, we first compare the diagonal square roots in Figure~\ref{fig:cov_compare_diag} and then the off-diagonals in Figure~\ref{fig:cov_compare_corr}. Figure~\ref{fig:cov_compare_diag} shows the ratio of the diagonal square roots of the data jackknife covariance matrix versus the mock-based covariance matrix. We see that the two diagonals are consistent in amplitude. There is some non-unity structure along the LoS direction at the smallest $r_p$ bin. This is likely attributed to the limited fidelity of the fiducial mock HOD, which we randomly picked from the initial MC chain at fiducial cosmology. Regardless, the point is that we find the overall amplitude of the two covariance diagonals to be consistent. 

Figure~\ref{fig:cov_compare_corr} compares the off-diagonal structures of the two covariance matrices, by showing their respective correlation matrices in the top left and bottom right triangles. The axis denote a flattened indexing of the $(r_p, \pi)$ bins, in a column-by-column order. The correlation function is simply the covariance matrix normalized by the diagonal, as defined in Equation~\ref{equ:corr}. 
By definition, the diagonal of the correlation matrix is 1. We see in Figure~\ref{fig:cov_compare_corr} that the two correlation matrices have very similar structure, with the mock-based correlation matrix showing higher signal-to-noise in the further off-diagonal terms. Combined with the fact that the two matrices have consistent diagonals, we conclude that the jackknife covariance matrix computed on CMASS data is consistent with a mock-based covariance matrix that does not suffer from limited sample volume.

\section{Phase correction}
\label{sec:phase}
Specifically, because our 85 cosmology boxes are run with the same initial condition phases, that means our emulator predictions are also at fixed phase. Thus the sample variance due to this fixed phase is frozen into our emulator prediction, and could potentially bias the emulator predictions relative to the data, which assumes an underlying ``true'' phase. However, we can remove this sample variance through a phase correction term to our model prediction in the form of 
\begin{equation}
    \xi^\mathrm{pred}_\mathrm{true}(r_p, \pi) = \xi^\mathrm{pred}_\mathrm{fixed}(r_p, \pi)[1+\delta(r_p, \pi)],
    \label{equ:phase_correct}
\end{equation}
where $\xi^\mathrm{pred}_\mathrm{true}(r_p, \pi)$ is the model prediction at true phase, and $\xi^\mathrm{pred}_\mathrm{fixed}(r_p, \pi)$ is the raw model prediction at fixed phase. $\delta(r_p, \pi)$ symbolizes a fractional correction term from fixed phase to true phase. In principle, this correction largely removes the effect of sample variance due to a limited volume with fixed phase. However, we do not quantify the reduction to the mock sample variance term here as it constitutes a rather minor contribution to the overall error budget. Nevertheless, this effect will become more important for upcoming DESI analyses as the data covariances become close to an order of magnitude smaller. 

\begin{figure}
    \centering
    \hspace*{-0.3cm}
    \includegraphics[width = 3.6in]{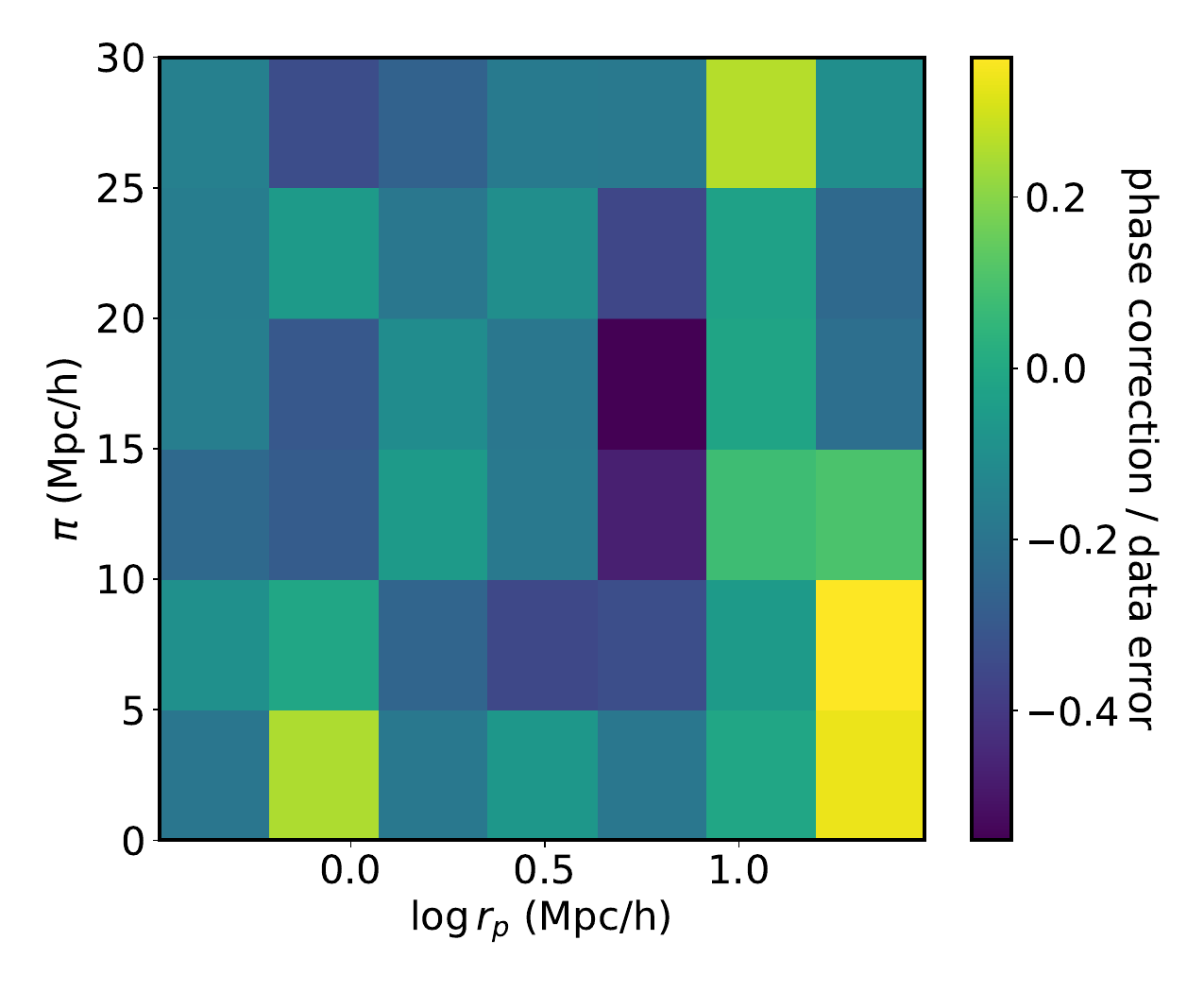}
    \vspace{-0.3cm}
    \caption{The phase correction term estimated at fiducial cosmology and averaged over 10 HODs, divided by the jackknife error bars from data. We see that the phase correction is a subdominant effect as its magnitude is largely within $30\%$ of the data uncertainty, with both a positive and negative spread. The median across separation bins is approximately $-15\%$, suggesting the overall correction is negative. }
    \label{fig:phase_correct}
\end{figure}

It is also the case that the phase correction term $\delta(r_p, \pi)$ should depend on both cosmology and HOD, but as a first order approximation, we assume such dependencies to be small compared to other uncertainties and only estimate the correction term at fixed cosmology and HOD. Specifically, we pick 10 HODs with high likelihood from the MC chain run at the c000 fiducial cosmology. We populate the 25 random phase boxes with these HODs, averaging over 16 realizations with different random seeds. We then estimate $\delta(r_p, \pi)$ by taking the fractional difference between the average $\xi$ prediction over the 10 HODs at all 25 boxes, and the average prediction at just ph000, which is our fixed phase. We illustrate this difference in Figure~\ref{fig:phase_correct}, compared to the data uncertainties.

We see that the phase correction is a subdominant effect compared to the data uncertainties as its magnitude is largely within $30\%$ of the data uncertainty. The overall correction is slightly negative with a median across separation bins of approximately $-15\%$. We do not see any significant structure as a function of separation. The fact that this correction term is sub-dominant supports our initial claim that its cosmology and HOD dependency is not significant for this analysis. In our subsequent analysis, we also find that the inclusion vs exclusion of this correction term does not significantly bias our inference.

\section{Robustness against priors}
\label{subsec:flat_prior}
While we adopt a covariant Gaussian prior for our fiducial analysis, we have also tested our constraints against different prior choices. First, we test different scaling to the size of the covariant Gaussian prior, inflating the prior until the posterior constraints fully converge. \red{We also test parameter recovery (Figure~\ref{fig:recovery}) with different inflated Gaussian width to ensure convergence.}

\begin{figure}
    \centering
    \hspace*{-0.3cm}
    \includegraphics[width = 3.4in]{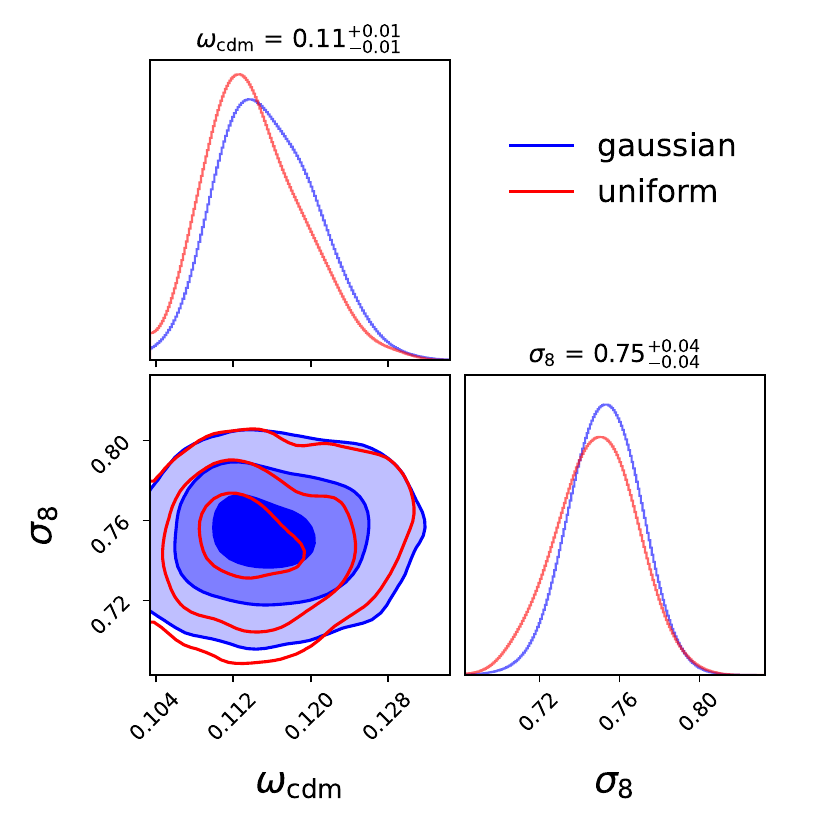}
    \vspace{-0.3cm}
    \caption{The constraints on $\sigma_8$ and $\omega_\mathrm{cdm}$ from Gaussian prior in blue and flat prior in red. The two sets of constraints are statistically consistent.}
    \label{fig:prior_test}
\end{figure}

\begin{figure*}
    \centering
    \hspace*{-0.3cm}
    \includegraphics[width = 7.5in]{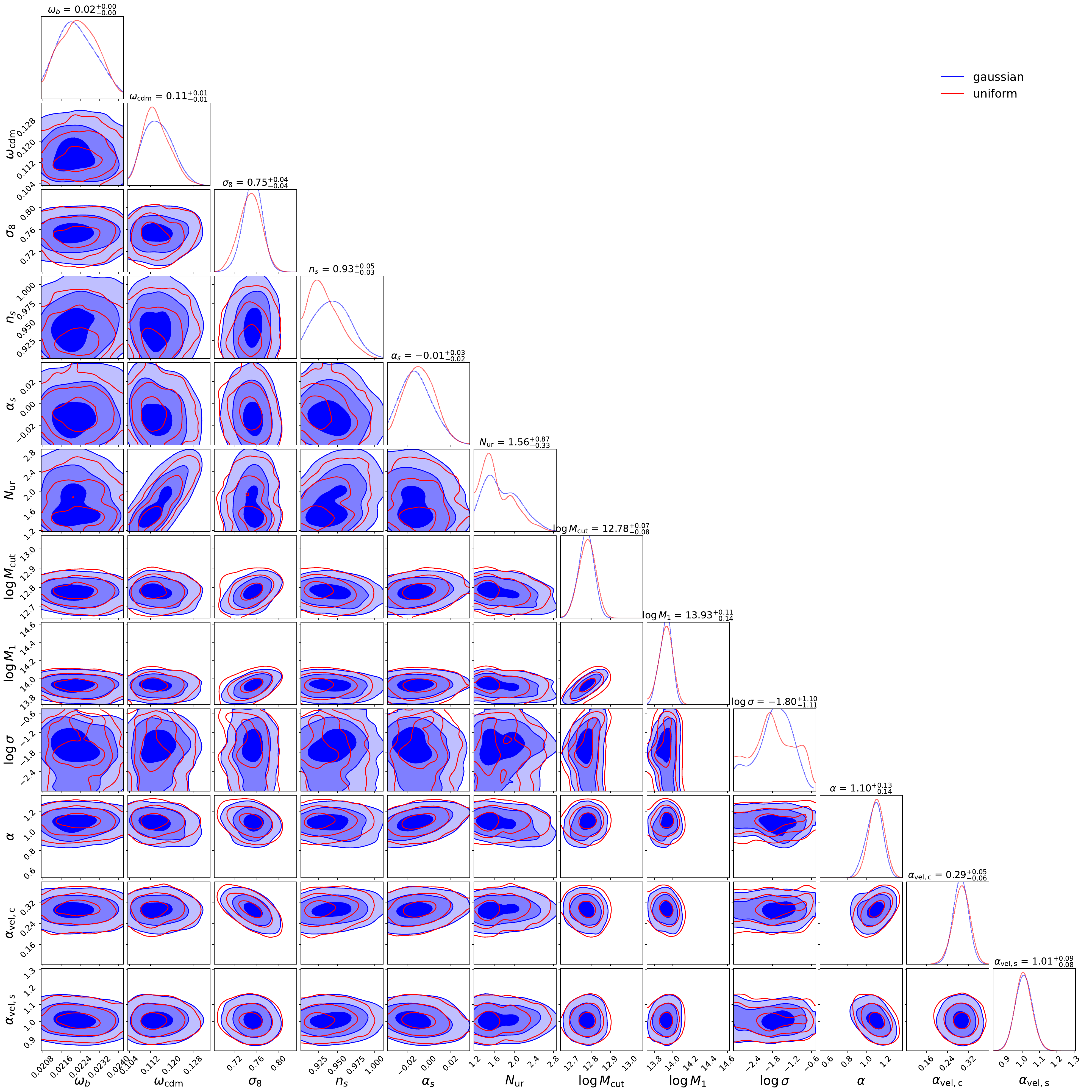}
    \vspace{-0.3cm}
    \caption{The full 2D marginalized constraints on cosmology and HOD parameters from the CMASS small-scale RSD measurement using flat priors (in red) and the fiducial Gaussian priors (in blue). The two sets of constraints are statistically consistent.}
    \label{fig:prior_test_full}
\end{figure*}

Here we repeat our fiducial analysis with a set of bounded flat priors instead of a Gaussian prior. To construct a flat prior, we identify the minimum bounding ellipse that encapsulates our full cosmology and HOD training sets, and we uniformly sample our prior from within that ellipse. After carrying out the the posterior sampling with \textsc{dynesty} on the same CMASS data, we find that the flat priors result in posterior constraints are statistically consistent with those from broad Gaussian priors. Specifically, we find less than $1\sigma$ deviations between the best-fit values of the two runs, and the posterior widths of the two fits also agree. \red{We show the full posterior constraints on the cosmology and HOD parameter space with flat priors and the fiducial Gaussian priors in Figure~\ref{fig:prior_test_full}. We see that the two sets of posterior constraints are consistent. Figure~\ref{fig:prior_test} zooms in on the constraints on $\sigma_8$ and $\omega_\mathrm{cdm}$ to show that the key cosmology results of this paper is robust to prior choices.} The two sets constraints are essentially identical in both best-fit values and in width. The title of the 1D marginalized PDFs show the $95\%$ constraints using flat priors. Besides the posterior constraints, the two runs also return consistent goodness-of-fits numbers. For the ``without environment-based bias'' case, the flat priors result in a best-fit $\log\mathcal{L} = -19.3$, compared to $\log\mathcal{L} = -18.9$ with broad Gaussian priors (Table~\ref{tab:fiducial_fits}). 

\section{\textsc{dynesty} vs MCMC}
\begin{figure*}
    \centering
    \hspace*{-0.3cm}
    \includegraphics[width = 6in]{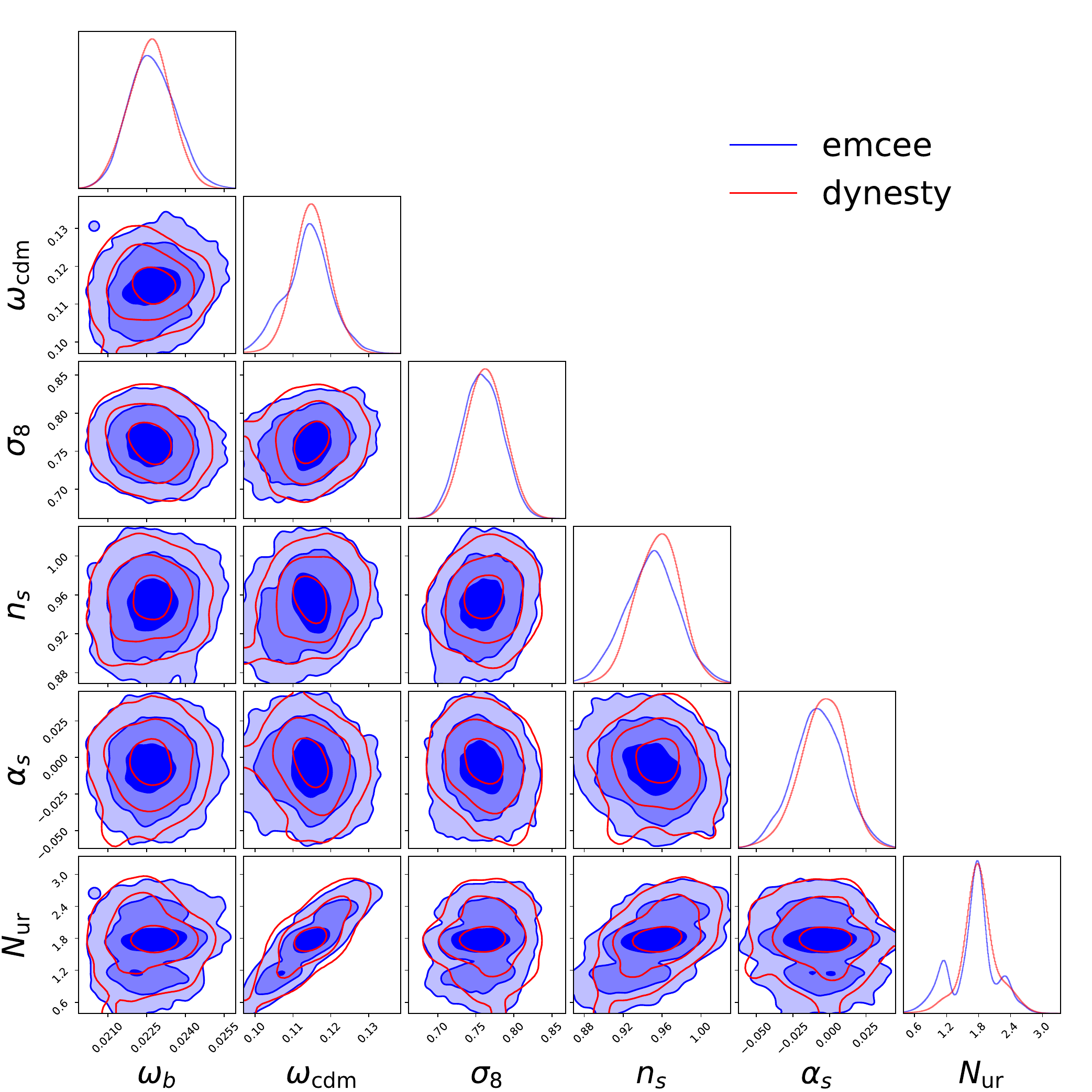}
    \vspace{-0.3cm}
    \caption{The fiducial posterior constraints on the cosmological parameters using two different sampling codes: \textsc{dynesty} in red and \textsc{emcee} in blue. While there is evidence that \textsc{dynesty} constraints tend to be a little tighter in certain parameters, the two sets of constraints are largely consistent.}
    \label{fig:emcee}
\end{figure*}

In this section, we test the robustness of our results against different samplers. Specifically, we test our fiducial nested sampler \textsc{dynesty} against a traditional MCMC sampler, \textsc{emcee} \citep[][]{2013Foreman}. We repeat our fiducial CMASS RSD analysis with \textsc{emcee}, using the same emulator and the same prior set up. In terms of MCMC choices, we use the affine-invariant ensemble sampler, 60 random walkers, and otherwise default settings. We burn in by running a set of short chains and then re-initializing the sampler with the high-likelihood points. Figure~\ref{fig:emcee} compares the cosmology posteriors of our fiducial analysis with \textsc{emcee} in blue and \textsc{dynesty} in red. First, we do not see any significant bias between the two samplers. Second, the constraints are largely consistent between the two samplers. There is evidence that \textsc{dynesty} tends to produce slightly tighter constraints under its default settings (Lange et al. in prep), therefore we value performing a cross-validation with additional sampler. However, we conclude that the difference is not significant for this analysis. 

\bsp	
\label{lastpage}
\end{document}